\documentclass[%
 reprint,
superscriptaddress,
 amsmath,amssymb,
 aps,
]{revtex4-2}

\usepackage{graphicx}
\usepackage{dcolumn}
\usepackage{bm}
\usepackage{hyperref}
\hypersetup{colorlinks,linkcolor={red},citecolor={blue},urlcolor={black}}

\begin{document}

\preprint{APS/123-QED}

\title{Effective rectification of THz electromagnetic fields in a ferrimagnetic iron garnet}

\author{T.G.H. Blank}
\affiliation{Radboud University, Institute for Molecules and Materials, 6525 AJ Nijmegen, the Netherlands.}
\affiliation{Department of Applied Physics, Eindhoven University of Technology, P.O. Box 513, Eindhoven 5600 MB, the Netherlands.}
\author{E.A. Mashkovich}
\affiliation{University of Cologne, Institute of Physics II, Cologne D-50937, Germany.}
\author{K.A. Grishunin}
\affiliation{Radboud University, Institute for Molecules and Materials, 6525 AJ Nijmegen, the Netherlands.}
\author{C. Schippers}
\affiliation{Department of Applied Physics, Eindhoven University of Technology, P.O. Box 513, Eindhoven 5600 MB, the Netherlands.}
\author{M.V. Logunov}
\affiliation{Kotel'nikov Institute of Radioengineering and Electronics, 125009 Moscow, Russia.}
\author{B. Koopmans}
\affiliation{Department of Applied Physics, Eindhoven University of Technology, P.O. Box 513, Eindhoven 5600 MB, the Netherlands.}
\author{A.K. Zvezdin}
\affiliation{Prokhorov General Physics Institute of the Russian Academy of Sciences, 119991 Moscow, Russia.}
\author{A.V. Kimel}
\affiliation{Radboud University, Institute for Molecules and Materials, 6525 AJ Nijmegen, the Netherlands.}

\date{\today}

\begin{abstract}
It is found that single-cycle THz electromagnetic fields efficiently excite a GHz spin resonance mode in ferrimagnetic Tm$_3$Fe$_5$O$_{12}$, despite the near absence of GHz spectral components in the exciting THz pulse. By analyzing how the efficiency of excitation depends on the orientation and strength of the THz electric field, we show that it can be explained in terms of the nonlinear THz inverse Cotton-Mouton effect. Here, the THz electric field gets effectively rectified and acts on the ferrimagnetic spins as a uni-polar effective magnetic field pulse. This interpretation is confirmed by a theoretical model based on the phenomenological analysis of the effective magnetic field, combined with the equations of motion derived from the effective Lagrangian for a ferrimagnet. Moreover, by using the outcome of two-dimensional THz spectroscopy, we conjecture a quantum-mechanical interpretation of the observed effect in terms of stimulated Raman scattering of THz photons by the crystal-field split f-f electronic transitions of Tm$^{3+}$. 
\end{abstract}

\maketitle

\section{\label{sec:Introduction}Introduction}
The development of ultrafast magnetism opened up a research field \cite{PhysRevLett.76.4250}, which explores new regimes of spin dynamics triggered in ferro, ferri, and antiferromagnetic materials by stimuli much shorter than the time required to reach thermodynamic equilibrium ($\sim 100$~ps). Consequently, in such regimes of spin dynamics, the conventional approximations of equilibrium thermodynamics to describe magnetic phenomena fail, and the resulting spin motion is often counter-intuitive. For instance, even though Curie’s principle \cite{curie1894symetrie} of equilibrium thermodynamics predicts that “the symmetry of the causes are to be seen in the effects” and thus solely heating cannot result in magnetization reversal, it was shown that ultrafast heating induced by femto or picosecond laser pulses or picosecond electrical pulses is able to reverse the magnetization in ferrimagnetic materials \cite{PhysRevLett.99.047601, Radu2011, Ostler2012, doi:10.1126/sciadv.1603117}. Similarly, in equilibrium thermodynamics, the fastest and the least dissipative route of magnetization reorientation seem to be always mutually excluding. However, this seemingly impossible exclusive combination was achieved by employing the effect of photo-induced magnetic anisotropy by a femtosecond laser-pulse, demonstrating the simultaneously record-fast and least-dissipative writing of a magnetic bit \cite{Stupakiewicz2017}. Naturally, this counter-intuitive but very appealing regime of spin dynamics attracted the attention of researchers in applied magnetism, including the fields of spintronics, magnonics, and magnetic data storage. This interest, in turn, fuels the search for ever new ultrashort stimuli to enable more ultrafast and even less dissipative writing of magnetic bits. 

While femto and picosecond laser pulses in the near-infrared and the visible spectral range are the most popular stimuli in ultrafast magnetism, it was realized that nearly single-cycle THz electromagnetic pulses \cite{Hebling:02}, consisting of photons with a thousand times smaller energy, can affect the spins in magnetic media in a more energy efficient non-thermal way. In particular, it was shown that the magnetic component of such THz pulses can directly couple to spins via Zeeman torque \cite{Kampfrath2011}. At the same time, such an approach is associated with a significant disadvantage. The time integral of the electromagnetic field of the THz pulses, as in the case of any other freely propagating electromagnetic wave in a neutral medium, is strictly zero and the net effect of such a stimulus on spins is thus questionable. But recently, it was shown that THz electric fields can be effectively rectified and thus become much more efficient in controlling spins \cite{nonspincontrol, Schlauderer2019}. Such rectification of THz fields has so far only been demonstrated in canted antiferromagnetic media \cite{PhysRevLett.123.157202}. Here, we fill the gap and demonstrate that a similar mechanism of effective field rectification can also be realized in ferrimagnets. 

In this article, we show that a single-cycle THz pulse in ferrimagnetic Tm$_3$Fe$_5$O$_{12}$ is able to excite not only a THz \cite{PhysRevLett.127.037203}, but also GHz mode of spin resonance. The excitation of the latter mode is surprising as its frequency is nearly absent in the spectrum of the THz pulse. By analyzing how the amplitude of the GHz mode depends on the orientation and the strength of the THz electromagnetic fields, we rule out heating and show that the fields are effectively rectified and that the GHz spin resonance is excited by an effective magnetic field generated in the medium due to the inverse Cotton-Mouton effect (ICME) \cite{ICMEobersvation, PhysRevB.97.224430}. These findings are supported by a phenomenological analysis of the rectified field in combination with a Lagrangian model of spin dynamics \cite{Davydova_2019}, which all fit the experimental observations. Quantum mechanically, the process can be described in terms of stimulated Raman scattering \cite{PhysRev.143.372}, where a first photon with frequency $\omega_1$ brings the electron to the excited state, while a second photon with frequency $\omega_2$ stimulates fast recombination to the Stokes-shifted ground state accompanied by an emission of a magnon with frequency $\omega = \omega_1 - \omega_2$. By employing two-dimensional (2D) THz spectroscopy \cite{Lu2018}, we show that the excited state 
is characterized by a lifetime shorter than $1$~ps. This fact practically excludes that the GHz mode is excited via the second THz mode of spin resonance.

The paper is organized as follows. Section~\ref {sec:Experimental} provides details about the sample and describes the experimental setup. Section~\ref{sec:Results} presents the main experimental findings. These findings are supported by the theory presented in section~\ref{sec:Theory}, where we employed a Lagrangian approach to describe the magnetization dynamics, driven by an effective rectified magnetic field due to the ICME. We suggest a possible microscopic mechanism of the latter in the discussion of section~\ref{sec:Discussion}, supported by additional measurements using 2D THz spectroscopy. Our conclusions are summarized in the final section~\ref{sec:Conclusions}.

\section{\label{sec:Experimental}Sample and Experimental Setup}
The particular ferrimagnet that we examined was a 19~$\mu$m thick film of bismuth and gallium substituted thulium iron garnet (TmIG) Tm$_{2}$BiFe$_{4.2}$Ga$_{0.8}$O$_{12}$ grown by liquid phase epitaxy on a $500$~$\mu$m gadolinium gallium garnet (Gd$_3$Fe$_5$O$_{12}$) substrate with $(111)$ orientation. The garnet structure has eight formula units per unit cell and has space-group symmetry Ia$\overline{3}$d (point group $O_h$) \cite{GELLER195730}. In the parent compound, Tm$_3$Fe$_5$O$_{12}$, two in five Fe$^{3+}$ ions can be found in octahedral surroundings of O$^{2-}$ and the other three in a tetrahedral environment \cite{wohlfarth1986handbook}. The spins of the tetrahedrally surrounded Fe$^{3+}$ ions couple antiferromagnetically to the spins of those in the octahedral environment. The two Fe$^{3+}$ magnetic sublattices have different magnetizations, this difference results in a net magnetization $\mathbf{M}_{\mathrm{Fe}}$. The spins of Tm$^{3+}$ couple antiferromagnetically with respect to $\mathbf{M}_{\mathrm{Fe}}$, resulting in a net magnetization $\mathbf{M}_{\mathrm{Tm}}$. Because the exchange interaction between the iron sublattices is large compared to any other exchange interaction between the three sublattices (Tm-Tm and Tm-Fe), it is sufficient to treat the two iron sublattices as one \cite{Levitin, neelferri}. Therefore, although Tm$_3$Fe$_5$O$_{12}$ is, in reality, a three-sublattice ferrimagnet, it is effectively treated as a more conventional two-sublattice Tm-Fe ferrimagnet with net magnetization $\mathbf{M} = \mathbf{M}_{\mathrm{Fe}} + \mathbf{M}_{\mathrm{Tm}}$ and N\'{e}el vector $\mathbf{L} = \mathbf{M}_{\mathrm{Fe}} - \mathbf{M}_{\mathrm{Tm}}$. 

In the garnet studied here, a part of the Tm$^{3+}$ ions were substituted by Bi$^{3+}$ to enhance the magneto-optical Faraday effect \cite{Hibiya_1985, Hansen, TmBiFeGaO12_substitutionsgerhard}. Moreover, some Fe$^{3+}$ ions on the tetrahedral sites \cite{PhysRev.148.522} were partly substituted with non-magnetic Ga$^{3+}$ to reduce $\mathbf{M}_{\mathrm{Fe}}$ and thus ensure that a possible ultrafast spin reorientation is not hampered by the need to facilitate an ultrafast exchange of large angular momentum  between the lattice and the spin system. Moreover, a sufficiently large Ga$^{3+}$ dilution should grant a magnetization compensation temperature to Tm$_3$Fe$_5$O$_{12}$ that usually has no compensation point due to the small magnetic moment of Tm$^{3+}$ \cite{Hansen}, but such a temperature was not observed in the present composition. Finally, both the parent compound as well as the substituted versions display uniaxial magnetic anisotropy along the out-of-plane $[111]$ axis \cite{Hansen}, as was confirmed by magneto-optical measurements \cite{PhysRevLett.127.037203}.

\begin{figure}[b!]
    \centering
    \includegraphics{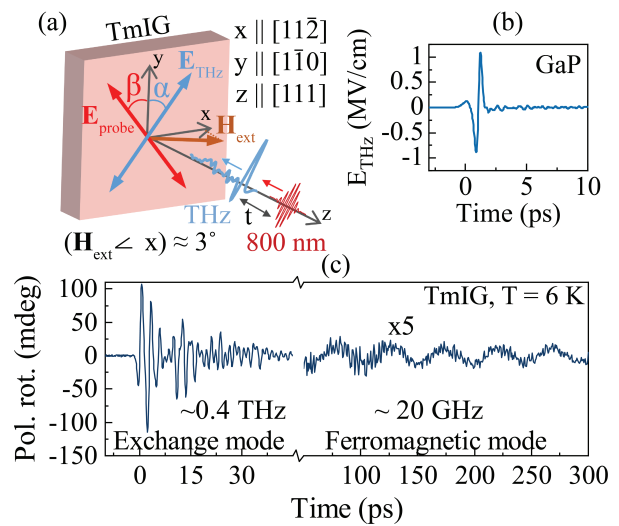}
    \caption{(a) Schematic illustration of the experimental scheme. (b) The calibrated waveform of the THz pulse measured by electro-optic sampling in GaP. The THz polarization is initially along the $y$-axis, but it could be rotated using wire-grid polarizers. (c) Typical THz-induced transient of the probe-polarization rotation, measured at $T = 6$~K with an applied external field of $110$~mT, pump polarization $\alpha = -45^\circ$ and probe polarization $\beta = 0^\circ$. The data at long timescales has been multiplied by a factor of $5$ for visibility. }
    \label{fig:setup}
\end{figure}

We performed a pump and probe experiment on the sample by employing single-cycle THz-pulses generated by tilted-pulse-front optical rectification in LiNbO$_3$ \cite{Hebling:02, Hirori_review_4ftheta}, yielding pulses with a peak electric field strength up to $\sim1$ MV/cm in focus as calibrated by electro-optic sampling in a slab of GaP-($110$) (see Fig.~\ref{fig:setup}(b)). The THz pulses were brought to temporal and spatial overlap with low-intensity near-infrared (NIR) probe pulses with a central wavelength of $800$~nm and pulse duration of $100$~fs. The experimental scheme and coordinate system are depicted schematically in Fig.~\ref{fig:setup}(a), indicating the THz pump and NIR probe polarization angles $\alpha$ and $\beta$ defined with respect to the $y$-axis. The polarizations were controlled by a set of wire-grid polarizers for the THz pump and a half wave-plate for the probe pulse. The probe pulse transmitted through the sample and its THz-induced rotation in polarization was mapped by a combination of a Wollaston prism and a set of balanced photodiodes. By measuring the polarization rotation as a function of the time-retardation $t$ between the pump and probe pulses, we traced the THz-induced spin dynamics. 

Our previous results showed that the magneto-optical signal of the transmitted probe pulse originates exclusively from out-of-plane magnetization components \cite{PhysRevLett.127.037203}. Therefore it can be expected that no dynamic magneto-optical signal due to magnetization precession will be detected when the equilibrium magnetization is out of the plane. To this end, a magnetic field was applied predominantly in the plane in order to saturate the magnetization in the plane, slightly titled at a small angle $\delta \sim 3^\circ$. The tilt was required to be able to excite THz spin dynamics for every polarization of the THz pulse (see Ref.~\cite{PhysRevLett.127.037203}), as is explained in more detail in the conclusions of Appendix~\ref{app:Lagrangian}. Finally, XRD analysis of the sample (see Supplemental Material~\footnote{\label{footnote}See the Supplemental Material for the XRD analysis of the sample, which includes Refs.~\cite{doi:10.1063/1.4929691, Ciubotariu2019}.}) confirmed its $[111]$ orientation. Moreover, the analysis provided us with the orientation of the crystallographic axes with respect to the experimental coordinate system $x \parallel [11\overline{2}]$ and $y \parallel [1\overline{1}0]$.

\section{\label{sec:Results}Results}
Figure~\ref{fig:setup}(c) shows a typical dynamical polarization rotation transient. In our previous article \cite{PhysRevLett.127.037203}, we showed that the ultrafast THz dynamics in the first $50$ picoseconds can be attributed to the ferrimagnetic THz Kaplan-Kittel exchange mode \cite{doi:10.1063/1.1699018}. Due to the unequal $g$-factors of the Tm and Fe sublattices, the Zeeman torque (or magnetic dipole interaction) acts differently on each sublattice, rendering a relatively efficient resonant excitation of the mode. Moreover, the bismuth-substitution yields a strong magneto-optical Faraday effect, which made detection with a good signal-to-noise ratio possible. In addition to the previously reported THz mode, the transients also reveal oscillations at a much lower (GHz) frequency. So low frequencies are typical for the ferromagnetic resonance (FMR) mode in a ferrimagnet, as was also predicted by the theory of Kaplan and Kittel. Given that the duration of our THz pulse $\sim 1$~ps is much shorter than the period of the mode $\sim 50$ ps, a resonant excitation by the THz magnetic field similar to what was seen with the exchange mode is unlikely. Instead, the excitation must be impulsive in nature. In order to reveal the excitation mechanism of this supposedly FMR mode, we measured the dynamics as a function of the pump polarization angle $\alpha$, the probe polarization angle $\beta$, the strength of the THz pump electric field, the strength of the external magnetic field as well as sample temperature.

\begin{figure}[t!]
    \centering
    \includegraphics{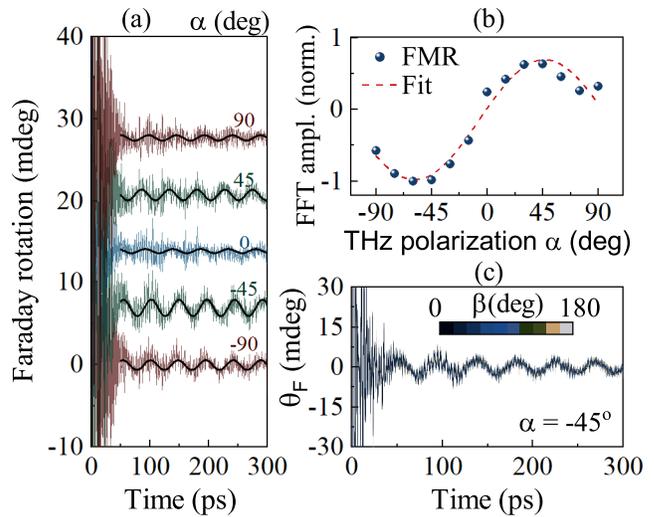}
    \caption{(a) THz induced dynamics as a function of the pump polarization angle $\alpha$, measured at a temperature of $6$~K and external field of $110$~mT. The data have been fitted by sine functions shown by the solid black lines. Two maxima at approximately $\pm 45^\circ$ can be observed, with nearly opposite phases. (b) Normalized peak Fourier amplitudes of the mode for different $\alpha$, where positive and negative values indicate the different phases of the dynamics. (c) THz-induced dynamics measured at the optimal THz polarization $\alpha = -45^\circ$ and field $B_{\mathrm{ext}} = 130$~mT, for various angles of the initial probe polarization $\beta$, showing no variations. }
    \label{fig:polscans}
\end{figure}

\subsection{Pump and probe polarization}
The measured dynamics are strongly dependent on the polarization angle $\alpha$ of the THz pump pulse (see Fig~\ref{fig:polscans}(a)), which rules out heating as the dominant mechanism. The optimal excitation occurs for $\alpha = \pm 45^\circ$ (see Fig.~\ref{fig:polscans}~(b)). The extrema are slightly asymmetric, such that the excitation is actually stronger for $\alpha = -45^\circ$. The dynamics for negative and positive $\alpha$ are approximately in opposite phases. These facts imply that the excitation mechanism is $\pi$-periodic with respect to $\alpha$. This is in contrast with the $2\pi$ periodicity of the excitation of the exchange mode due to the Zeeman torque \cite{PhysRevLett.127.037203}. Instead, the results are more comparable to the nonlinear excitation of the quasi-ferromagnetic mode in antiferromagnetic FeBO$_3$, where the largest signals were also found when the electric (or magnetic) field of the pump pulses was polarized at $\pm 45^\circ$ from the net magnetization \cite{PhysRevLett.123.157202}.  

\begin{figure}[t!]
    \centering
    \includegraphics{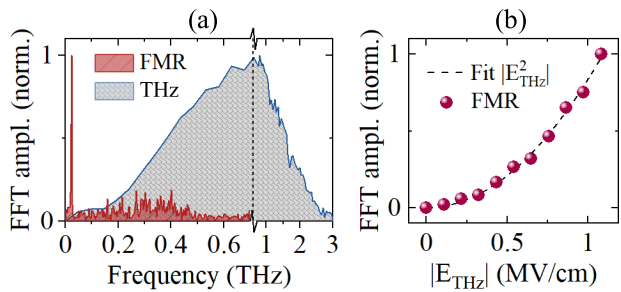}
    \caption{(a) Typical Fourier spectrum of the signal (red) and the exciting THz pulse (blue). The sharp peak in the signal spectrum that corresponds to the FMR mode falls out of the THz spectrum. (b) Peak FFT amplitude of the FMR mode as a function of THz electric field amplitude, measured at a temperature of $6$~K, THz polarization $\alpha = -45^\circ$ and $B_{\mathrm{ext}} = 110$~mT. When fitting the data with a power law dependence of the form $E_{\mathrm{THz}}^\gamma$ for variable $\gamma$, the data is best fitted with a quadratic dependence $\gamma = 2$.}
    \label{fig:powerscan}
\end{figure}

\begin{figure*}[t!]
    \centering
    \includegraphics{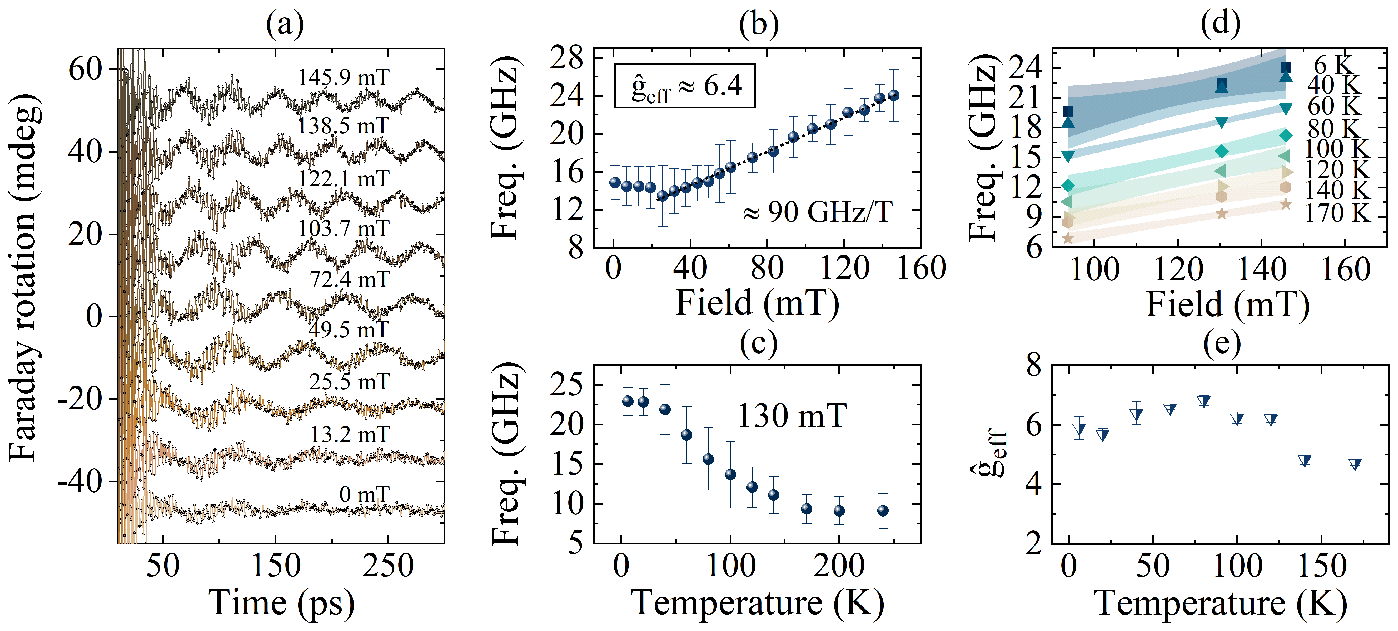}
    \caption{(a) THz-induced transients for several strengths of the applied magnetic field, measured at $T = 6$~K and $\alpha = -45^\circ$. Part (b) shows the extracted frequencies, where the width of the error bars equals the FWHM of the fitted Gaussian in the FFT spectrum. The slope of the linear part of the curve provides a reasonable estimate $\hat{g}_{\mathrm{eff}}$ for the effective $g$-factor $g_{\mathrm{eff}}$. (c) Extracted frequency as a function of temperature for a fixed magnetic field of $130$~mT, where the bars again depict the FWHM. (d) The frequency measured at three different external magnetic fields, for various temperatures, where the blurred areas depict the $95\%$ confidence bands of the fitted line. (e) The resulting estimations of the effective $g$-factor based on the fitted slope in (d), the bars denote the standard error.}
    \label{fig:fieldscans}
\end{figure*}

Next, it can be seen in Fig.~\ref{fig:polscans}(c) that the signal does not depend on the initial probe polarization angle $\beta$. This implies that the observed dynamics are a result of the magneto-optical Faraday effect, to which only the out-of-plane magnetization contributes $\theta_{\mathrm{F}}(t) \propto M_z(t)$ \cite{PhysRevLett.127.037203}. Therefore, starting from Fig.~\ref{fig:polscans}, we referred to the induced polarization rotation as Faraday rotation $\theta_{\mathrm{F}}(t)$.

\subsection{THz amplitude}
Figure~\ref{fig:powerscan} shows the FFT spectrum of the dynamics (red) measured at the optimal THz polarization angle $\alpha = -45^\circ$, as well as the spectrum of the THz pump-pulse (blue). The sharp red peak that corresponds to the GHz (FMR) mode lies completely out of the spectrum of the exciting THz pulse. This fact indicates that the excitation is most likely nonlinear. This nonlinearity was confirmed by measuring the peak Fourier amplitude of the mode as a function of the strength of the THz electric field $E_{\mathrm{THz}}$ (see Fig.~\ref{fig:powerscan}(b)). The data could only be fitted accurately with a quadratic dependence $E_{\mathrm{THz}}^2$. 

\subsection{External magnetic field and temperature} The assignment of the exact origin of the FMR mode is not straightforward, because TmIG is, in reality, a three-sublattice ferrimagnet. This means that possible spin resonances are associated not only with the spins of Fe$^{3+}$ but also with those of the Tm$^{3+}$ ions. To reveal the origin of the mode, we studied the frequency as a function of the external magnetic field and temperature. Figure~\ref{fig:fieldscans}(a) shows that small changes in the external magnetic field have an enormous impact on the frequency of the mode. It can be seen in Fig.~\ref{fig:fieldscans}(b) that the frequency initially decays, but for larger fields $> 25$~mT it increases approximately linearly. Theoretically, the linear slope can be related to the $g$-factor of the FMR mode \cite{gurevich1996magnetization}. In the case of FMR involving only iron, which has a $g$-factor of $g \approx 2$, the slope should amount to $28$~GHz/T. However, in our case, the slope is approximately $\sim 90$~GHz/T, which allows us to give an estimate of the \textit{effective $g$-factor} $g_{\mathrm{eff}}$ is about $\hat{g}_{\mathrm{eff}} \approx 6.4$. 
\begin{figure}[b!]
    \centering
    \includegraphics{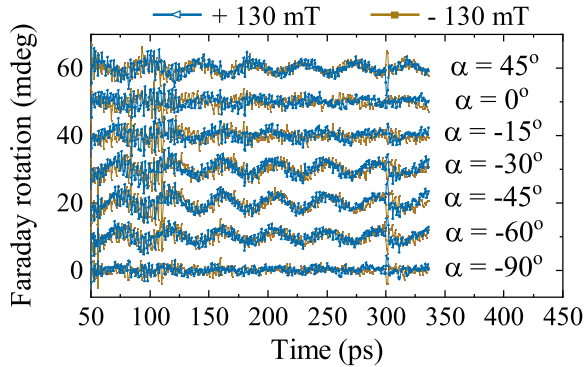}
    \caption{THz-induced dynamics measured at $T = 6$~K, for several $\alpha$ and two opposite field polarities $B_{\mathrm{ext}} = \pm 130$~mT. For the angles $\alpha$ where the excitation is strong, the phase of the mode remains unaltered under field reversal. However, it appears that the dynamics measured around the minimum $\alpha = 0^\circ$ is affected by the polarity of the field.}
    \label{fig:pmB}
\end{figure}

The fact that the effective $g$-factor is so much larger than that of iron, tells us that the thulium sublattice must be involved. Let $g_{\mathrm{Tm}}$ and $g_{\mathrm{Fe}}$ be the $g$-factors of the individual Tm and Fe sublattices, respectively. The theory of Kittel predicts that the effective $g$-factor, in that case, is given by \cite{PhysRev.115.1587}:
\begin{equation}
\label{geff}
    g_{\mathrm{eff}}(T) =\frac{M_{\mathrm{Fe}}(T) - M_{\mathrm{Tm}}(T)}{ g_{\mathrm{Fe}}^{-1}M_{\mathrm{Fe}}(T) - g_{\mathrm{Tm}}^{-1}M_{\mathrm{Tm}}(T)}.
\end{equation}
The effective $g$-factor can thus have a strong temperature dependence in the vicinity of the angular momentum compensation point, i.e. when  $g_{\mathrm{Fe}}^{-1}M_{\mathrm{Fe}}(T) = g_{\mathrm{Tm}}^{-1}M_{\mathrm{Tm}}(T)$. Otherwise, no strong temperature dependence of the effective $g$-factor is expected. Here we should note that the magnetization compensation point in this sample either does not exist or is slightly above $0$~K.

Figure \ref{fig:fieldscans}(c) shows that the frequency at a fixed magnetic field does drop significantly when increasing the temperature. But at the same time, it can be seen in Fig.~\ref{fig:fieldscans}(d) that the slope of the frequency as a function of the field does not decrease as a function of temperature. It means that the decay in frequency observed in Fig.~\ref{fig:fieldscans}(c) is a result of a temperature-dependent effective field such as magnetic anisotropy, and not of a decreasing effective $g$-factor. In Fig.~\ref{fig:fieldscans}(e), we show the effective $g$-factors estimated from the slopes of the linear fits in Fig.~\ref{fig:fieldscans}(d). It reveals that the effective $g$-factor has a peculiar temperature dependence and remains to be large. Large values of the effective $g$-factor have also been observed in a Tm$_{3}$Fe$_5$O$_{12}$ compound with La, Ca, and Ge substitutions \cite{doi:10.1143/JPSJ.43.705}. Here, the result could be fitted by assuming that the Tm angular momentum is quenched, i.e. $g_{\mathrm{Tm}} = 0 $. It is therefore likely that the peculiar temperature dependence of $g_{\mathrm{eff}}$ is greatly influenced by the dilution of Bi and Ga. Other factors might also play a role, such as the potential noncollinear arrangement of the Tm spins \cite{DOROSHEV1994290, TCHEOU19701751} that were observed in these materials, which raises interesting questions for further studies. In any case, the fact that the $g$-factor remains well above that of the free electron ($g=2$) in the entire temperature range from $6$ to $170$~K implies that the thulium sublattice is involved in the FMR mode.

Finally, we measured how the dynamics depend on a change in the polarity of the external magnetic field. Figure~\ref{fig:pmB} shows that for the excitation maxima $\alpha = \pm 45^\circ$, the phase of the dynamics is independent of the change of field polarity $\pm \textbf{B}_{\mathrm{ext}}$. This result is in contrast with the excitation of the THz mode by Zeeman torque reported in Ref.~\cite{PhysRevLett.127.037203}, where the detected magneto-optical transients did reverse their phase upon a change of polarity of the external magnetic field. This fact again confirms that the dominating mechanism of excitation of the FMR mode in the studied TmIG is not due to the linear coupling of the THz magnetic field to the spins via the Zeeman torque. Only around $\alpha = 0^\circ$, one can distinguish a very weak signal that changes sign upon the field reversal. This signal could be interpreted to be due to the linear excitation mechanism, but it is clear that this mechanism is quite inefficient and negligible in comparison with nonlinear excitation.

To summarize, the obtained experimental dependencies reveal that THz electromagnetic pulse excites GHz oscillations of the magneto-optical Faraday effect in thulium iron garnet, which probes the out-of-plane projection of the magnetization dynamics $M_z(t)$. The excitation is $\pi$-periodic with respect to the THz pump polarization $\alpha$ and reaches a maximal efficiency for $\alpha = \pm 45^\circ$. The dependency of the amplitude of the oscillations clearly reveals that the mechanism of the excitation is nonlinear with respect to the strength of the THz field. The dependency of the frequency of the oscillations on the applied external magnetic field implies that the oscillations must be assigned to a spin resonance in the compound. The unusually large effective $g$-factor deduced from the measurements and the typical GHz frequency suggests that the oscillations are associated with the low-frequency FMR mode in the system of two macro spins formed by the magnetizations of iron $\mathbf{M}_{\mathrm{Fe}}$ and thulium $\mathbf{M}_{\mathrm{Tm}}$ sublattices. To further support this interpretation, we propose a theoretical model described in the next section.

\section{\label{sec:Theory}Theory}This section is structured as follows: in subsection~\ref{sec:FMRfreq} we derived an expression for the FMR frequency. In the derivation, we employed an effective ferrimagnetic Lagrangian that was simplified to describe dynamics corresponding to the FMR mode only. A complete analysis with the total effective Lagrangian is presented in Appendix~\ref{app:Lagrangian}. From the derivations, it becomes clear that the spin dynamics can be triggered if THz light acts as an effectively rectified magnetic field. Hence, in subsection~\ref{sec:EnergyICME}, we phenomenologically describe the rectification in terms of the inverse Cotton-Mouton effect. We derived the form of the effective rectified magnetic field based on point-group symmetry. Finally, in subsection~\ref{sec:GeneralLagrangian}, we combine the Lagrangian with the ICME interaction potential to obtain equations for the THz-induced motion of the spins.  
\subsection{\label{sec:FMRfreq}Frequency of FMR}
Let $\theta, \varphi$ be the polar and azimuthal angles of the net magnetization $\mathbf{M} = \mathbf{M}_{\mathrm{Fe}} + \mathbf{M}_{\mathrm{Tm}} \equiv m(\sin\theta\cos\varphi, \sin\theta\sin\varphi, \cos\theta)$, where $m = M_\mathrm{Fe} - M_\mathrm{Tm}$. The total effective Lagrangian density describing magnetization dynamics of a two-sublattice ferrimagnet is given in Appendix \ref{app:Lagrangian}. The equations of motion derived from this Lagrangian possess two eigenmodes, corresponding to the THz exchange mode and the GHz FMR mode \cite{PhysRevLett.127.037203}. When considering only the FMR mode, the description can be simplified using the fact that the involved magnetizations remain antiparallel during the dynamics \cite{gurevich1996magnetization}, while for the exchange mode, the sublattices become mutually canted. The way to impose on the Lagrangian that we only want to consider the FMR mode is, therefore, to enforce the sublattices to remain antiparallel. This can be done by letting the antiferromagnetic exchange coupling parameter $\lambda$, which defines the exchange energy $U_{\mathrm{ex}} = -\lambda \mathbf{M}_{\mathrm{Fe}} \cdot \mathbf{M}_{\mathrm{Tm}}$, go to infinity $\lambda \to -\infty$. This \textit{ferromagnetic approximation} does not influence the FMR mode and its frequency (see Appendix~\ref{app:Lagrangian}). In this case, the effective Lagrangian reduces to the low-frequency effective Lagrangian density
$\mathcal{L}_{\mathrm{FM}}$:
\begin{equation}
    \label{eq:Lagrangian}
    \mathcal{L}_{\mathrm{FM}} = - \frac{m}{\gamma_{\mathrm{eff}}} \dot\varphi \cos\theta  - U(\theta, \varphi).
 \end{equation}
Here, $\gamma_{\mathrm{eff}}$ is the effective gyromagnetic ratio $\gamma_{\mathrm{eff}} \equiv g_{\mathrm{eff}} \mu_B /\hbar $ and $U(\theta,\varphi)$ is the static potential energy density, which contains uniaxial anisotropy and Zeeman interaction with the external magnetic field:
\begin{equation}
    \label{eq:U}
    U(\theta, \varphi) = -K_u \frac{(\mathbf{M} \cdot \hat{\mathbf{z}})^2}{m^2} - \mathbf{M}\cdot\mathbf{H}_{\mathrm{ext}},
\end{equation}
where $K_u>0$ is the uniaxial anisotropy constant and $\mathbf{H}_{\mathrm{ext}} = (H_x, 0, H_z)$ the external magnetic field (in Tesla). Similar to the experiment, we let the external magnetic field be slightly tilted away from the sample plane, i.e. $H_z/H_x = \tan\delta$ 
(see Fig.~\ref{fig:setup}). The ground state angles $\theta_0$, $\varphi_0$ can be found by minimization of \eqref{eq:U}. Afterwards, the equations of motion that are given by the Euler-Lagrange equations can be linearized around the ground-state angles $\theta = \theta_0 + \theta_l$, $\varphi = \varphi_0 + \varphi_l$ with $\theta_l, \varphi_l \ll 1$:
\begin{equation}
\begin{aligned}
    \label{EulerLagrange}
    0 = \frac{d}{dt}\frac{\partial\mathcal{L}}{\partial\dot\theta} - \frac{\partial\mathcal{L}}{\partial\theta} \approx -\frac{m}{\gamma_{\mathrm{eff}}} \dot\varphi_l \sin\theta_0 + U''_\theta(\theta_0, \varphi_0) \theta_l, \\
    0 = \frac{d}{dt}\frac{\partial\mathcal{L}}{\partial\dot\varphi} - \frac{\partial\mathcal{L}}{\partial\varphi} \approx \frac{m}{\gamma_{\mathrm{eff}}} \dot\theta_l \sin\theta_0 + U''_\varphi(\theta_0, \varphi_0)\varphi_l.
\end{aligned}
\end{equation} 
where we introduced the notation $U''_{\theta}(\theta_0, \varphi_0) \equiv \tfrac{\partial^2}{\partial\theta^2}U(\theta,\varphi)\big\rvert_{\theta = \theta_0, \varphi = \varphi_0}$. The FMR frequency is given by the eigenfrequency of these coupled equations, which we separately derived for a purely in-plane field and tilted field configuration.

\subsubsection{In-plane field ($\delta = 0^\circ$)}
\begin{figure}[b!]
    \centering
    \includegraphics{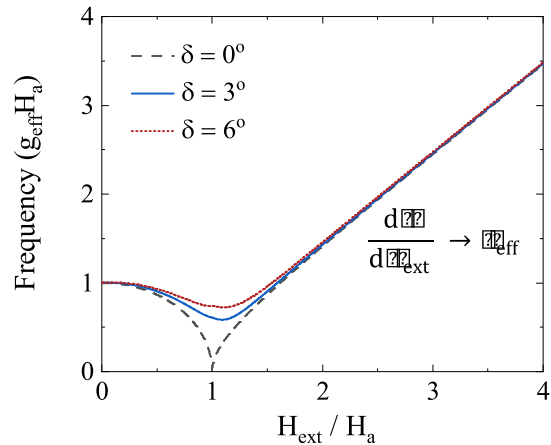}
    \caption{Theoretical curves for the FMR frequency as a function of the external magnetic field, for different tilting angles $\delta$.}
    \label{fig:theoryfrequency}
\end{figure}

When ignoring the tilting of the external magnetic field, the static energy potential becomes:
\begin{equation}
    \label{eq:U_inplane}
    U(\theta, \varphi) = -K_u\cos^2\theta - m H_x \sin\theta\cos\varphi.
\end{equation}
Minimization w.r.t. $\varphi$ yields $\varphi_0 = 0$ for $H_x >0$ and $\varphi_0 = \pi$ for $H_x <0$. For the minimization w.r.t. $\theta$, we need to consider two regimes: when the applied external magnetic field is greater or smaller than the anisotropy field $H_a \equiv 2K_u/m$:
\begin{equation}
    \label{eq:theta_0field}
    \theta_0 = \begin{cases}
    \sin^{-1}{\frac{H_x}{H_a}} &\text{for $ |{H_x}| < H_a$,}\\
    \pi/2  & \text{for $  |{H_x}| \geq H_a$.}\end{cases}
\end{equation}
Using these ground-state angles, the eigenfrequency of the equations of motion \eqref{EulerLagrange} can be found:
\begin{equation}
    \label{FM_0field}
    \omega_{\mathrm{FM}} = \begin{cases}
    \gamma_{\mathrm{eff}}\sqrt{H_a^2 - H_x^2} &\text{for $ |{H_x}| < H_a$,}\\
    \gamma_{\mathrm{eff}}\sqrt{H_x(H_x-H_a)}  & \text{for $  |{H_x}| \geq H_a$.}\end{cases}
\end{equation}
The solution for this case ($\delta = 0^\circ$) is plotted in Fig.~\ref{fig:theoryfrequency}, which shows that the frequency drops to zero when $|{\mathbf{H}_{\mathrm{ext}}}| \to H_a$, and afterward approaches a linear trend. We do observe the linear increase of the frequency in the experiment, but we do not see such a significant drop at low fields. Therefore, we need to include the small tilt of the external magnetic field $\delta \neq 0^\circ$.
\subsubsection{Tilted field ($\delta \neq 0^\circ$)}
In the case of a tilted magnetic field, the static potential energy is given by:
\begin{equation}
    \label{Uangle}
    U(\theta, \varphi) = - K_u \cos^2\theta - mH_x \sin\theta\cos\varphi - mH_z\cos\theta.
\end{equation}
Again we have that $\varphi_0 = 0$ for $H_x > 0$ and $\varphi_0 = \pi$ for $H_x < 0$, which means $H_x\cos\varphi = |{H_x}|$ and we can minimize \eqref{Uangle} only with respect to $\theta$:
\begin{equation}
    \label{eq:U'}
    \frac{1}{m} U'_\theta(\theta) = \cos\theta(H_a\sin\theta - |{H_x}|) + H_z \sin\theta = 0.
\end{equation}
To be able to solve this equation, we treat the out-of-plane field as a perturbation with respect to the case of an in-plane field. Therefore, we substitute $\theta = \theta_0 - \epsilon $ in Eq.~\eqref{eq:U'} with $\theta_0$ as in the in-plane case (Eq.~\eqref{eq:theta_0field}) and assumed $\epsilon \ll 1$. We retained maximally cubic terms $\sim \epsilon^3$ and solved the third-degree polynomial for $\epsilon(H_x, H_z)$. The solutions gives the new ground-state angle $\theta_0' = \theta_0 - \epsilon(H_x,H_z)$. From here, we again linearized the Euler-Lagrange equations~\eqref{EulerLagrange} and calculated expressions for the FMR resonance mode as presented in Eq.~\eqref{FM_cantedfield}.   
\begin{widetext}
\begin{equation}
    \label{FM_cantedfield}
    \omega_{\mathrm{FM}} = \begin{cases}
   \gamma_{\mathrm{eff}}\frac{\sqrt{H_a^2 - H_x^2 + \left(\tfrac{7}{2}H_x^2 - 2H_a^2\right)\epsilon^2 + \left(3H_x(H_a^2-H_x^2)^\frac{1}{2} + H_zH_x\right)\epsilon + H_z(H_a^2 - H_x^2)^\frac{1}{2}}}{\sqrt{1 - \epsilon(H^2_a/H^2_x - 1)^\frac{1}{2} - \tfrac{1}{2}\epsilon^2}} &\text{for $ |{H_x}| < H_a$,}  \\
   \\
   \frac{\gamma_{\mathrm{eff}}}{(1-\tfrac{1}{2}\epsilon^2)^{\frac{1}{2}}}\sqrt{H_x\left[\left(H_x-H_a\right) + H_z\epsilon + (2 H_a - \tfrac{1}{2}H_x)\epsilon^2\right] }  & \text{for $  |{H_x}| \geq H_a$.}
   \end{cases}
\end{equation}
\end{widetext}
We solved the third-degree polynomial of $\epsilon$ numerically for different external magnetic fields and tilt angles and calculated the corresponding resonance frequencies. The result is plotted in Fig.~\ref{fig:theoryfrequency}, which shows that we can obtain a better agreement to our data of Fig.~\ref{fig:fieldscans} when accounting for the tilted field. We also see that the minimal frequency occurs when the external magnetic field equals the uniaxial anisotropy field. Looking back at experimental data from Fig.~\ref{fig:fieldscans}(b), we see that the dip occurs at approximately $25$~mT which therefore allows us to estimate the size of the uniaxial anisotropy field $H_a \approx 25$~mT. However, when inserting this value into Eq.~\ref{FM_cantedfield}, our theory predicts $\omega_{\mathrm{FM}}=\gamma_{\mathrm{eff}} H_a$ at zero external field which amounts to a frequency of $(2\pi)^{-1}\omega_{\mathrm{FM}} \approx 2.24 $~GHz, while experimentally we observe $\sim 14$~GHz (see Fig.~\ref{fig:fieldscans}(b)). This unaccounted frequency offset of about $ 12$~GHz suggests that we should include some effective biasing field with a magnitude of about $130$~mT. At this point, we have no definite answer to where this field stems from. Shape anisotropy should play a role given the clear domain patterns seen in this sample without an external magnetic field \cite{PhysRevLett.123.157202}. Alternatively, the so-called ``double umbrella structure'' - a noncollinear arrangement of the thulium ions seen at helium temperatures in TmIG using neutron diffraction \cite{TCHEOU19701751, DOROSHEV1994290} - might be involved. Also, strain in the sample induced by the Gd$_3$Fe$_5$O$_{12}$ substrate could be the source of the missing field \cite{Ciubotariu2019}. Fortunately, this open question does not obstruct the theoretical treatment of the rectified effective magnetic field by the ICME which will be held in the coming section.

\subsection{\label{sec:EnergyICME}Energy considerations for the ICME}
In this section, we derive an effective interaction potential starting from the most basic principles of light-matter interaction. This interaction potential will then enter the Lagrangian in the next section. When an oscillating electric field of light $\mathbf{E}(t)$ enters a non-absorbing medium, it interacts with the medium by inducing electric polarization. The change in interaction energy density $dW$ due to the increase of electric polarization $\mathrm{d}\mathbf{P}$ is given by $dW = - \mathbf{E}\cdot \mathrm{d}\mathbf{P}$. Here we ignored higher-order multipole contributions as well as magnetic dipole interactions because they are expected to be weak. In the linear optical approximation, the amount of induced polarization is linear to the applied electric field $P_i = \chi_{ij} \epsilon_0 E_j$, with $\epsilon_0$ the dielectric permittivity of vacuum and $\chi_{ij}$ the electric susceptibility tensor. Therefore, the total interaction potential energy density $W$, after integration, becomes:
\begin{equation}
    \label{interactionpotential}
    W = -\frac{1}{2}\chi_{ij} \epsilon_0 E_i E_j.
\end{equation}
The electric susceptibility tensor is related to the dielectric permittivity $\epsilon_{ij} = (1+\chi_{ij})\epsilon_0$. The presence of static magnetization induces magneto-optical birefringence in the medium, and modifies this dielectric permittivity \cite{zvezdin1997modern}:
\begin{equation}
    \label{deps_main}
    \delta\epsilon_{ij} = k_{ijk}M_k + \xi_{ijkl}M_kM_l + \cdots,
\end{equation}
where the third-rank antisymmetric axial tensor $k_{ijk}$ and fourth-rank symmetric polar $g_{ijkl}$ tensor describe the well-known \textit{Faraday and Cotton-Mouton effects}, respectively. The relevant interaction potential that describes the interaction of light and magnetization can then be found by substituting $\epsilon_0(\delta \chi_{ij}) =  (\delta \epsilon_{ij})$ from Eq.~\eqref{deps_main} in \eqref{interactionpotential} to obtain \cite{nonspincontrol}:
\begin{equation}
    \label{W}
    W = -\frac{1}{2}k_{ijk}E_iE_jM_k - \frac{1}{2}\xi_{ijkl}E_iE_jM_kM_l.
\end{equation}
On one hand, magnetization may therefore induce birefringence in the medium and modify the properties of light via the well-known Faraday and Cotton-Mouton effects. On the other hand, inversely, the presence of light may induce magnetization. That is, by thermodynamics, the electric field of light acts as an \textit{effective magnetic field} $\mathbf{H}_{\mathrm{eff}} = - \frac{\delta W}{\delta \mathbf{M}}$ \cite{RevModPhys.82.2731}. The appearance of an effective field originating from the two terms in Eq.~\eqref{W} are known as the inverse Faraday effect (IFE) and ICME, respectively.

To obtain an expression for the effective fields, we make a Fourier expansion of the wave and assume it is monochromatic (the extension to non-monochromatic light is straightforward) $\mathbf{E}(t) = \mathrm{Re}[\bm{\mathcal{E}}(\omega) \exp(-i\omega t)] = \tfrac{1}{2}\left[\bm{\mathcal{E}}(\omega) \exp(-i\omega t) +\bm{\mathcal{E}}^*(\omega) \exp(i\omega t) \right] $, where $\bm{\mathcal{E}}(\omega)$ is the ``Jones vector''. The Fourier components of the polarization are again related to those of the electric field $P_i(\omega) = \epsilon_0\tilde{\chi}_{ij}(\omega)  \mathcal{E}_j(\omega)$ where $\tilde{\chi}_{ij}(\omega)$ is the optical susceptibility. Because we consider a non-absorbing medium, the tensor $\tilde{\chi}_{ij}(\omega)$ is required to be Hermitian $\tilde{\chi}^*_{ij}(\omega) =\tilde{\chi}_{ji}(\omega)$ \cite{LandauLifshitz}. Using this fact, the interaction potential \eqref{interactionpotential} becomes \cite{pitaevskii1961electric}:
\begin{equation}
    \label{interactionpotential}
    W = -\frac{1}{4}\epsilon_0 \tilde{\chi}_{ij} \mathcal{E}^*_i\mathcal{E}_j + \mathit{h.f.},
\end{equation}
where we neglect the high-frequency (\textit{h.f.}) terms as they average out on the relevant timescales \cite{LandauLifshitz, PhysRev.130.919}. Similarly, Eq.~\eqref{W} becomes:
\begin{equation}
     W = -\frac{1}{4}k_{ijk}\mathcal{E}^*_i\mathcal{E}_jM_k - \frac{1}{4}\xi_{ijkl}\mathcal{E}^*_i\mathcal{E}_jM_kM_l 
\end{equation}By Neumann's principle \cite{Neumann}, the tensors $k_{ijk}$ and $\xi_{ijkl}$ should be invariant under the crystallographic point group operations. In an isotropic or cubic medium, $k_{ijk}$ is an antisymmetric imaginary tensor with only a single nonzero tensor component $k_{xyz} = k_{zxy} = k_{yzx} = -k_{xzy} = -k_{yxz} = -k_{zyx} = -i k$ \cite{PhysRev.143.574, PhysRev.130.919, birss1964symmetry}, and the corresponding effective field is:
\begin{eqnarray}
\mathbf{H}_{\mathrm{IFE}} = \frac{ik}{4}\bm{\mathcal{E}}(\omega)\times \bm{\mathcal{E}}^*(\omega).
\end{eqnarray}
Therefore, a circularly polarized light pulse generates a rectified effective magnetic field with opposite directions for left/right-handed polarized light \cite{Nat435Kimel, PhysRev.130.919, PhysRevLett.15.190,PhysRev.143.574}.
However, for linearly polarized light, the effective field is zero. Given that our THz pulses are linearly polarized, we only consider the inverse Cotton-Mouton field \cite{PhysRevB.97.224430}:
\begin{equation}
    \label{eq:HICME}
    (\mathbf{H}_{\mathrm{ICME}})_l = \frac{\xi_{ijkl}}{2}\mathcal{E}^*_i(\omega)\mathcal{E}_j(\omega)M_k.
\end{equation}
Using the crystallographic point group $O_h$, we can find the minimal expression for the tensor $\hat{\xi}$. The nonzero tensor components in the standard cubic coordinate system $x\parallel [100]$, $y\parallel[010]$, $z\parallel [001]$ are tabulated in Ref.~\cite{birss1964symmetry}, where it can be seen that the tensor has only two independent components $\xi_{xxxx}$ and $\xi_{xxyy}$. We transformed the tensor $\xi_{ijkl}$ to our experimental coordinate system (see Fig.~\ref{fig:setup}), where $z\parallel [111]$, $x\parallel [11\overline{2}]$ and $y\parallel [1\overline{1}0]$. In this coordinate system, the tensor can be expressed in Voight notation by a $6\times6$ matrix $\tilde{\xi}$:
\begin{equation}
    \tilde{\xi} = \begin{pmatrix}
3\xi_1 & \xi_1 & \xi_1 + \xi_2 & 0 & -\sqrt{2}\xi_2 & 0\\
\xi_1 & 3\xi_1 & \xi_1 + \xi_2 & 0& \sqrt{2}\xi_2 & 0 \\
\xi_1 + \xi_2 & \xi_1+\xi_2 & 3\xi_1 - \xi_2 & 0 & 0 & 0 \\
0 & 0 & 0 & \xi_1 + \xi_2 & 0 & \sqrt{2}\xi_2\\
-\sqrt{2}\xi_2 & \sqrt{2}\xi_2 & 0 & 0 & \xi_1+\xi_2 & 0 \\ 
0 & 0 & 0 & \sqrt{2}\xi_2 & 0 & \xi_1
\nonumber
\end{pmatrix}
\end{equation}
where $\xi_1 \equiv \frac{\xi_{xxxx}}{6} + \frac{\xi_{xxyy}}{2}$ and $\xi_2 =\frac{\xi_{xxxx}}{6} - \frac{\xi_{xxyy}}{2}$. This gives the ICME interaction energy from Eq.~\eqref{W}:
\begin{equation}
    \label{W2}
    \begin{aligned}
    W_{\mathrm{ICME}} = -\frac{\xi_1}{2}\left(2(E_xM_x + E_y M_y)^2 + E_{\mathrm{THz}}^2m^2\right) \\ 
    - \frac{\xi_2}{2}\Big(E_{\mathrm{THz}}^2 M_z^2 + 2\sqrt{2}M_x M_z(E_y^2 - E_x^2) \\ + 4\sqrt{2}E_xE_yM_yM_z\Big).
    \end{aligned}
\end{equation}
where we used that $m= \sum_i\sqrt{M_i^2}$ constant. Note that the latter acts as a holonomic constraint on the system, making the expression for the ICME field as in Eq.~\eqref{eq:HICME} a bit naive because it presumes no constraints on the variables $M_i$. Although such a constraint complicates a Newtonian approach to describe the influence of $\mathbf{H}_{\mathrm{ICME}}$ on $\mathbf{M}$, in the Lagrangian approach  treated in the next subsection the problem is naturally circumvented. 

The Newtonian approach is anyhow insightful and we briefly treat it here before going back to the Lagrangian approach. We can obtain the approximate ICME field by assuming that $|H_{\mathrm{ext}}| \geq H_a$ while ignoring the tilt $M_z \ll M_x$ such that $\mathbf{M} = M_x\hat{\mathbf{x}}$. In that case, $M_x$ can be regarded as constant (only changing sign for $\pm H_{\mathrm{ext}}$), while $M_{y,z}$ are two independent variables. Furthermore, the THz electric field $\mathbf{E}_{\mathrm{THz}}(t)$ lies in the $xy$ plane, and therefore $\mathbf{E}_{\mathrm{THz}}(t) = E_{\mathrm{THz}}(t)(\sin\alpha, \cos\alpha, 0)$. Then, an approximate expression for the inverse Cotton-Mouton field derived from Eq.~\eqref{W} is given by:
\begin{equation}
    \mathbf{H}_{\mathrm{ICME}} = E_{\mathrm{THz}}^2 M_x \begin{pmatrix} 0\\ \xi_1\sin2\alpha  \\ \sqrt{2}\xi_2 \cos2\alpha  \label{ICMEfield}\end{pmatrix}.
\end{equation}
Given that the THz pulse duration of about $t_{\mathrm{THz}}\sim 1$~ps is much shorter than that of the period of the FMR mode, we could treat the effect of this field on magnetization $\mathbf{M}$ as an instantaneous (impulsive) torque $\mathbf{\tau} = \mathbf{M}\times\mathbf{H}_{\mathrm{ICME}}$, which triggers dynamics that can be described using the Landau-Lifshitz equation $\frac{\mathrm{d}\mathbf{M}}{\mathrm{d}t} = - \gamma \mathbf{M}\times\mathbf{H}$, with the initial condition $\mathbf{M}(t = 0) = (M_x, 0,0)$:
\begin{equation}
    \dot{\mathbf{M}}(t = 0) =  \gamma M_x^2 \begin{pmatrix} 0  \\ \sqrt{2}\xi_2 \cos2\alpha  \\ -\xi_1 \sin2\alpha. \label{ICMEfield}\end{pmatrix} \int_0^{t_{\mathrm{THz}}}\left[\frac{\mathrm{d}}{\mathrm{d}t}E^2_{\mathrm{THz}}\right] \, \mathrm{d}t,
\end{equation}
where $E_\mathrm{THz}(t)$ is the THz pulse electric field. It can be easily seen that the torque fulfills the quadratic dependence with respect to the THz field amplitude $\sim E_{\mathrm{THz}}^2$. Moreover, in order to have maximal torque at $\alpha = \pm 45^\circ$, we find out that $\xi_1$ must be dominant over $\xi_2$, i.e. $\xi_1 \ll \xi_2$. At the same time, it can be shown that the latter is also required for the invariance of the phase of the observed dynamics under field reversal as we showed experimentally in Fig.~\ref{fig:pmB}. This can be derived from the fact that we are only sensitive to the $z$-projection of the magnetization, while the precession of the magnetization around the external magnetic field is always right-handed. If the term $\xi_2$ would have been dominant, magnetization precession would have been launched with a $z-$projection that has an exactly opposite phase for $\pm \mathbf{H}_{\mathrm{ext}}$, unlike the experiment. This invariance of the phase on field polarity when $\xi_1 \ll \xi_2$ can be seen directly in the coming section after we solved the Lagrangian equations of motion. 

\subsection{\label{sec:GeneralLagrangian}Lagrangian equations of motion driven by the ICME}
To include light-matter interaction in the Lagrangian, we introduce the interaction energy $f(\theta,\varphi)$:
\begin{equation}
    \label{eq:Lagrangian}
    \mathcal{L}_{\mathrm{FM}} = - \frac{m}{\gamma_{\mathrm{eff}}} \cos\theta \dot\varphi - U(\theta, \varphi) + f(\theta,\varphi) ,
 \end{equation}
which contains Zeeman interaction with a dynamic THz magnetic field $\mathbf{h}(t)$ as well as the ICME of a THz electric field $\mathbf{E}(t)$ that is derived from the interaction potential in Eq.~\eqref{W2}:
\begin{equation}
\begin{split}
    \label{f(theta,varphi)}
    &f(\theta, \varphi) \quad =  \quad m\sin\theta\left(h_x\cos\varphi + h_y\sin\varphi\right) \\& +  m^2\xi_1\sin^2\theta\left[E_x^2\cos^2\varphi + 2E_xE_y\cos\varphi\sin\varphi + E_y^2\sin^2\varphi\right] \\& +\frac{m^2\xi_2}{2}\Big[E^2\cos^2\theta + 2\sqrt{2}\sin\theta\cos\theta\cos\varphi\left(E_y^2-E_x^2\right) \\& \qquad \qquad \qquad + 4\sqrt{2}E_xE_y\sin\theta\cos\theta\sin\varphi\Big].
\end{split}
\raisetag{1\baselineskip}
\end{equation}
Note that for the full Lagrangian (see Appendix~\ref{app:Lagrangian}), the driving energy contains several other terms that are pertinent to ferrimagnets (and absent for ferromagnets). The equations of motion derived from the Euler-Lagrange equations, after linearization, are given by:
\begin{equation}
    \begin{aligned}
    \label{eqmo1}
    U''_{\theta}(\theta_0, \varphi_0)\theta_l - \sin\theta_0\frac{m}{\gamma_{\mathrm{eff}}} \dot\varphi_l &=& f_\theta(t), \\
    U''_{\varphi}(\theta_0, \varphi_0)\varphi_l + \sin\theta_0 \frac{m}{\gamma_{\mathrm{eff}}}  \dot\theta_l &=& f_\varphi(t),
    \end{aligned}
\end{equation}
where the generalized force terms $f_\nu$ are defined for $\nu = \theta, \varphi$ and are evaluated around the ground-state angles:
\begin{equation}
    \label{formuladrivingterms}
    f_\nu(t) = - \left[\frac{\mathrm{d}}{\mathrm{d}t}\left(\frac{\partial f}{\partial \dot\nu}\right) - \frac{\partial f}{\partial \nu}\right]\Bigg\rvert_{\nu=\nu_0} = \frac{\partial f}{\partial \nu}\Big\rvert_{\nu = \nu_0}, 
\end{equation}
These generalized forces can be interpreted as torque. When ignoring the small canting angle $\delta$, we obtain the following driving terms evaluated for the case $|H_x|\geq H_a$:
\begin{equation}
    \begin{aligned}
     f_\theta(t) = -\sqrt{2} m^2\xi_2E_{\mathrm{THz}}^2 \cos\varphi_0 \cos2\alpha,  \quad \quad  \\
     f_\varphi(t) = m h_y\cos\varphi_0 + m^2\xi_1 E_{\mathrm{THz}}^2\cos^2\varphi_0 \sin 2\alpha .
     \end{aligned}
\end{equation}
The driving term $f_\theta(t)$ proportional to $\xi_2$ is maximal when $\alpha = 0^\circ, 90^\circ$, in contrast with the experiment. Therefore, we again conclude that $\xi_2 \ll \xi_1$ and essentially ignore this term. Next, as expected, it can be seen in $f_\varphi(t)$ that the THz magnetic field component $h_y(t)$, which is perpendicular to the equilibrium magnetization, could drive the precession. However, this term scales linearly with the THz amplitude, which is also not the dominant mechanism in the experiment. Theoretically, this term is also negligible as we will treat the excitation as instantaneous in which such linear driving terms disappear because $\int_0^{t_{\mathrm{THz}}} h(t) \mathrm{d}t = 0$. Therefore, we only consider the ICME term in $f_\varphi$ that is proportional to $\xi_1$. 

Experimentally, we are only sensitive to the $z$-component of the magnetization $M_z(t) = m\cos\theta(t) \approx -m \theta_l(t)$ (for $|{H_x}| \geq H_a$ in which case $\theta_0 = \pi/2$). In other words, we only detect dynamics of $\theta_l(t)$. The equation of motion for $\theta_l(t)$, found by differentiation and mutual substitution of Eqns.~\eqref{eqmo1}, is given by: 
\begin{equation}
    \ddot\theta_l + \omega_{\mathrm{FM}}^2 \theta_l = m \gamma_{\mathrm{eff}}\xi_1 \left[\frac{\mathrm{d}}{\mathrm{d}t}E_{\mathrm{THz}}^2\right]\cos^2\varphi_0 \sin 2\alpha.
\end{equation}
We treat the excitation as instantaneous (see ``photonic impact'' \cite{Zvezdin2020}), in which case the excited dynamics is fully determined by the initial condition $\theta_l(0) = 0$ and:
\begin{equation}
\dot\theta_l(0) = m \gamma_{\mathrm{eff}}\xi_1\cos^2\varphi_0 \sin 2\alpha \int_0^{t_{\mathrm{THz}}}\left[\frac{\mathrm{d}}{\mathrm{d}t}E_{\mathrm{THz}}^2\right]\,\mathrm{d}t.
\end{equation}
As $\varphi_0 = 0$ for $H_x > 0$ and $\varphi_0 = \pi$ for $H_x < 0$, we have that $\cos^2\varphi_0 = 1$ in both cases such that a change in field polarity does not change the phase of the dynamical out-of-plane magnetization $\dot{M}_z(0) \approx -m\dot\theta_l(0)$, just as we observed in the experiment (see Fig.~\ref{fig:pmB}). Moreover, the excitation is maximal and has a mutually opposite sign for $\alpha = \pm 45^\circ$, and scales quadratically with the THz field. Therefore, all experimentally observed features regarding the excitation are captured by this Lagrangian approach. 
\section{\label{sec:Discussion}2D spectroscopy and Discussion}
The predictions made by the phenomenological treatment of the ICME fit the experimental data very well, but it does not yet clarify the underlying microscopic mechanism. The effect is conventionally seen as a consequence of impulsive stimulated Raman scattering (ISRS) involving either resonant or off-resonant excitation of a certain electronic transition \cite{PhysRev.143.372, PhysRev.143.574, doi:10.1063/1.453733, PhysRevLett.99.167205, PhysRevB.77.094426, PhysRevB.78.104301, PhysRevResearch.2.043035}. It is thus interesting if we are able to assign a certain transition that mediates the ISRS process. Here, it is important to take into account the possibility of the recently discovered mechanism of magnonic Raman scattering~\cite{blank2022empowering}, where the GHz q-FM resonance in FeBO$_3$ was excited through resonant Raman scattering of the q-AF THz resonance. A notable similarity between FeBO$_3$ and the current experiment is that the maxima of excitation both occur when the THz polarization is $\alpha = \pm 45^\circ$ from the net magnetization. But at the same time, the materials are largely different since FeBO$_3$ is an antiferromagnet with Néel vector $\mathbf{L} \perp \mathbf{M}$, while collinear ferrimagnets such as TmIG have $\mathbf{L} \parallel \mathbf{M}$. In any case, if the exchange mode mediates the excitation of the FMR mode through magnonic Raman scattering, it should become visible in 2D THz spectroscopy \cite{blank2022empowering}. In this technique, instead of only one THz pulse, two THz pulses are applied at a mutual delay $\tau$. This allows us to track the Faraday rotation signal of the probe pulse when both THz pulses are present $\theta_{\mathrm{F},12}(t,\tau)$, but also the signals $\theta_{\mathrm{F},1}(t,\tau)$ and $\theta_{\mathrm{F},2}(t,\tau)$ obtained when only one of the two pulses excited the sample. From here, we obtained the nonlinear signal:
\begin{equation}
    \label{eq:thetaNL}
    \theta_{\mathrm{F, NL}}(t,\tau) \equiv \theta_{\mathrm{F},12}(t,\tau) - \theta_{\mathrm{F},1}(t,\tau) - \theta_{\mathrm{F},2}(t,\tau).
\end{equation}
\begin{figure}[b!]
    \centering
    \includegraphics{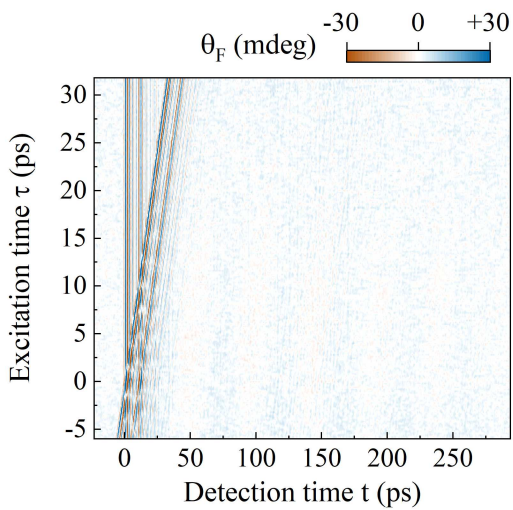}
    \caption{Result of $\theta_{\mathrm{F},12}(t,\tau)$ measured in 2D THz spectroscopy. Fast modulations shortly after the pulse arrival associated with the exchange mode can be observed, as well as faint contours which correspond to the FMR mode at longer timescales. However, no modulations in the excitation efficiency as a function of $\tau$ of the latter mode were found.}
    \label{fig:2D}
\end{figure}
If the exchange mode mediates the excitation of the FMR mode, this should become apparent as periodicity in the excitation efficiency of the FMR mode with a period defined by the frequency of the exchange mode. We performed this experiment and the result of $\theta_{\mathrm{F},12}(t,\tau)$ is shown in Fig.~\ref{fig:2D}. The measured signal does not show any periodicity of the mode, and neither did we observe a signal in the nonlinear part calculated with Eq.~\eqref{eq:thetaNL}. The only logical explanation for the absence of a nonlinear signal in 2D spectroscopy is that the nonlinear excitation is mediated by a state with a short lifetime $<1$~ps. In this case, no information will be carried over between two subsequent THz pulses which are separated by a time longer than their own pulse duration of $\sim 1$~ps. In other words, the second pulse delayed at a time $\tau >1$~ps will never be able to experience the presence of the first THz pulse, meaning that superposition holds and no nonlinear signal (see Eq.~\ref{eq:thetaNL}) will come up. 
\begin{figure}[t!]
    \centering
    \includegraphics{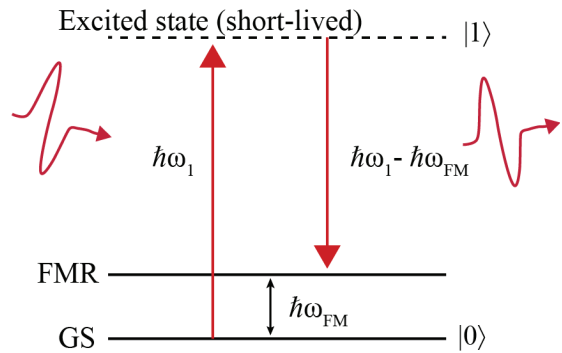}
    \caption{Schematic illustration of ISRS from a short-lived electronic excited state, which is believed to be the microscopic mechanism for the ICME.}
    \label{fig:ISRS}
\end{figure}
Therefore, the THz exchange mode does not mediate the ISRS mechanism and is not responsible for the excitation of the FMR mode. 

Figure \ref{fig:ISRS} illustrates the mechanism of ISRS via a short-lived excited state. The process could occur completely off-resonantly via electronic or phononic transitions which lie beyond the THz excitation spectrum, in which case the excited state is virtual. However, a resonant excitation via a state that lies within the THz pulse spectrum is expected to be more likely. The only candidates for such an excited state are the crystal-field split transitions of the Tm$^{3+}$ ground state. The dodecahedral crystal-field environment of the Tm\textsuperscript{3+} (4f\textsuperscript{12}) ions, with local symmetry described by the dihedral point-group D\textsubscript{2}, splits its ground-state multiplet \textsuperscript{3}H\textsubscript{6} ($J=6$) into $2J+1 = 13$ Stark levels. The energy level diagram for substituted Tm\textsuperscript{3+} ions in dodecahedral D\textsubscript{2} sites in Y\textsubscript{3}Al\textsubscript{5}O\textsubscript{12} was earlier obtained from experimental absorption and emission spectra in \cite{PhysRevB.40.9464, Tiseanu_1995} and more recently it was studied in \cite{doi:10.1021/acs.inorgchem.8b02009}. The lowest electric-dipole-allowed transition has an energy of 27~cm\textsuperscript{-1} corresponding to about $0.809$ THz. This energy is in the vicinity of a spectral feature we actually observed in this material \cite{PhysRevLett.127.037203}, but which we couldn't assign. Such a crystal-field split transition can be very short-lived and effectively act as an effective virtual electronic state from which light can scatter. Moreover, since there are no other transitions in our spectrum, we conjecture by deduction that ISRS of the lowest crystal-field split state of thulium is the responsible microscopic mechanism of excitation (see Fig.~\ref{fig:ISRS} for a schematic illustration of ISRS). In order to substantiate this hypothesis, one must develop a microscopic theory that takes into account both the Tm-Fe exchange interaction and the electronic structure of Tm$^{3+}$. This problem is beyond the scope of this article. 

\section{\label{sec:Conclusions}Conclusion}
We showed that a single-cycle THz pulse is able to excite a GHz magnon in ferrimagnetic TmIG. The experimental dependencies reveal that the excitation is a result of ICME, where the THz field of light becomes effectively rectified to generate a unipolar magnetic field pulse. The results are supported by the equations of motion obtained from an effective ferrimagnetic Lagrangian, using a phenomenological expression for the rectified field. We discussed the possible microscopic picture of ICME by considering the mechanism of ISRS. 2D spectroscopy ruled out magnon-magnon scattering, similar to what occurred in FeBO$_3$ \cite{blank2022empowering}, to be responsible. Instead, we conjectured that the effect is more similar to that in TmFeO$_3$ \cite{nonspincontrol}, and is based on light-induced scattering from the crystal-field split electronic states of Tm$^{3+}$. 

Our results demonstrate that nonlinear THz optomagnetic effects do no not only play a role in antiferromagnets but also in ferrimagnetic materials. In general, this nonlinearity facilitates a channel of energy transfer from the electric field of light to the magnetic spin system. Such a channel recently enabled coherent steering of spins over a potential barrier in antiferromagnetic TmFeO$_3$ \cite{Schlauderer2019}. Similarly, our results, therefore, open a way for future data-writing, spintronics, and magnonics applications based on ferrimagnets.  
\begin{acknowledgments}
The authors thank Sergey Semin and Chris Berkhout for their technical support. The work was supported by the Dutch Research Council (NWO). The authors declare that this work has been published as a result of peer-to-peer scientific collaboration between researchers. The provided affiliations represent the actual addresses of the authors in agreement with their digital identifier (ORCID) and cannot be considered as a formal collaboration between the aforementioned institutions.
\end{acknowledgments}

\appendix 
\begin{widetext}
\section{\label{app:Lagrangian}Total Effective Lagrangian}
The total effective Lagrangian density of a two-sublattice ferrimagnet in the vicinity of the compensation temperature can be written in terms of the polar and azimuthal angles of the net magnetization $\mathbf{M} = \mathbf{M}_{\mathrm{Fe}} + \mathbf{M}_{\mathrm{Tm}} \equiv m(\sin\theta\cos\varphi, \sin\theta\sin\varphi, \cos\theta)$ \cite{Davydova_2019, PhysRevLett.127.037203}:
    \begin{equation}
    \mathcal{L}_\mathrm{eff} = \frac{\chi_\perp}{2}\left[\left(\frac{\dot\theta}{\overline{\gamma}} + H_x\sin\varphi\right)^2 + \left(\left(\frac{\dot\varphi}{\overline\gamma}-H_z\right)\sin\theta + H_x\cos\theta\cos\varphi\right)^2\right] - \frac{m\dot\varphi}{\gamma_{\mathrm{eff}}}\cos\theta - U(\theta, \varphi) + \bar{f}(\theta,\varphi),
    \end{equation}
where $\chi_\perp \equiv 1/|{\lambda}|$ with $\lambda<0$ the exchange coupling parameter that defines the exchange energy density $U_{\mathrm{ex}} = - \lambda \mathbf{M}_{\mathrm{Fe}}\cdot\mathbf{M}_{\mathrm{Tm}}$, and $\overline\gamma \equiv \left(M_{\mathrm{Fe}} + M_{\mathrm{Tm}}\right)/\left(M_{\mathrm{Fe}}/\gamma_{\mathrm{Fe}} + M_{\mathrm{Tm}}/\gamma_{\mathrm{Tm}}\right)$. The static potential $U$ contains Zeeman interaction with the external magnetic field and magnetic anisotropy as defined in Eq.~\eqref{eq:U}. The interaction energy density $\bar{f}(\theta, \varphi)$ is equal to:
    \begin{equation}
\begin{aligned}
    \label{barf(theta,varphi)}
    \bar{f}(\theta, \varphi) = \frac{\chi_\perp}{2}\Bigg[\cos^2\theta(h_x\cos\varphi + h_y\sin\varphi)^2 + 2\left(\left(\frac{\dot\varphi}{\overline\gamma} - H_z\right)\sin\theta + H_x\cos\theta\cos\varphi\right)\cos\theta\left(h_x\cos\varphi + h_y\sin\varphi\right)\\ +2\left(\frac{\dot\theta}{\overline\gamma} + H_x\sin\varphi\right)\left(h_x\sin\varphi - h_y\cos\varphi\right) + \left(h_x\sin\varphi - h_y\cos\varphi\right)^2\Bigg]  + m\sin\theta\left(h_x\cos\varphi + h_y\sin\varphi\right)  \\+  m^2\xi_1\sin^2\theta\left[E_x^2\cos^2\varphi + 2E_xE_y\cos\varphi\sin\varphi + E_y^2\sin^2\varphi\right] \\ + \frac{m^2\xi_2}{2}\left[E^2\cos^2\theta + 2\sqrt{2}\sin\theta\cos\theta\cos\varphi\left(E_y^2-E_x^2\right) + 4\sqrt{2}E_xE_y\sin\theta\cos\theta\sin\varphi\right]
\end{aligned}
\end{equation}
Damping can be included through the Rayleigh function $\mathcal{R} = \frac{\alpha \mathcal{M}}{2\overline{\gamma}}\left(\dot\theta + \sin^2\theta\dot\varphi\right)$,
where $\mathcal{M} \equiv M_{\mathrm{Fe}} + M_{\mathrm{Tm}}$.
The general equations of motion are defined by the Euler-Lagrange equations:
\begin{equation}
\begin{aligned}
    \label{EulerLagrange2}
    \frac{d}{dt}\frac{\partial\mathcal{L}_\mathrm{eff}}{\partial\dot\theta} - \frac{\partial\mathcal{L}_\mathrm{eff}}{\partial\theta}  + \frac{\partial\mathcal{R}} {\partial \dot\theta} = 0, \quad \quad \quad \frac{d}{dt}\frac{\partial\mathcal{L}_\mathrm{eff}}{\partial\dot\varphi} - \frac{\partial\mathcal{L}_\mathrm{eff}}{\partial\varphi}  + \frac{\partial\mathcal{R}}{\partial \dot\varphi} = 0.
\end{aligned}
\end{equation}

The result can be linearized around the ground-state angles $\theta = \theta_0 + \theta_l$, $\varphi = \varphi_0 + \varphi_l$, $\theta_l, \varphi_l \ll 1$ (where $\theta_0$, $\varphi_0$ are as in Sec~\ref{sec:FMRfreq}). This gives the linearized equations of motion:
\newline
\begin{equation}
\begin{aligned}
\label{eqmotion}
\ddot\theta_l + \zeta \dot\theta_l + \frac{\overline{\gamma}^2}{\chi_\perp}U''_{\theta}(\theta_0)\theta_l - \sin\theta_0\frac{\overline\gamma^2}{\gamma_{\mathrm{eff}}} |{\lambda}| m\dot\varphi_l =\frac{\overline\gamma^2}{\chi_\perp} \bar{f}_\theta(t), \quad \quad \quad 
\ddot\varphi_l + \zeta\dot\varphi_l + \frac{\overline\gamma^2U''_{\varphi}(\theta_0, \varphi_0)}{\chi_\perp\sin\theta_0}\varphi_l + \frac{\overline{\gamma}^2}{\gamma_{\mathrm{eff}}}  |{\lambda}| m \dot\theta_l = \frac{\overline\gamma^2}{\chi_\perp}\frac{\bar{f}_\varphi(t)}{\sin\theta_0},
\end{aligned}
\end{equation}
where $\zeta = \tfrac{\alpha\bar{\gamma} \mathcal{M}}{\chi_\perp}$ and the driving terms can be calculated from the interaction energy $\bar{f}$ from \eqref{barf(theta,varphi)} using equation \eqref{formuladrivingterms}.
These equations have two eigenfrequencies, one corresponding to the exchange mode $\omega_{\mathrm{ex}} \approx  |{\lambda}|  \left(\gamma_{\mathrm{Tm}} M_{\mathrm{Fe}} - \gamma_{\mathrm{Fe}} M_{\mathrm{Tm}}\right)$,
and one corresponding to the FMR mode whose frequency is given by Eq.~\eqref{FM_cantedfield}.
The driving terms after linearization, taking only the leading contributions within the first order of $\epsilon$ into account, are given by:
\begin{equation}
\begin{aligned}
    \label{eq:drivingforces}
    \bar{f}_\theta(t) = \chi_\perp \overline\gamma^{-1} \dot h_y\cos\varphi_0 + \sqrt{2} m^2\xi_2 \cos\varphi_0 \left(E_x^2 - E_y^2\right) + \epsilon\left(  mh_x\cos\varphi_0  + 2m^2\xi_1E_x^2 \cos^2\varphi_0 - m^2\xi_2E(t)^2 \right) \\
     \bar{f}_\varphi(t) =   \left( m h_y\cos\varphi_0 +2 m^2\xi_1 E_xE_y\cos^2\varphi_0\right) + \epsilon\left(-\chi_\perp \overline\gamma^{-1} \dot h_x \cos\varphi_0 + 2\sqrt{2} m^2\xi_2E_xE_y\cos\varphi_0 \right)
\end{aligned}
\end{equation}
where we also ignored the terms $- \epsilon \chi_\perp \cos^2\varphi_0(h_x^2 + 2H_xh_x)$ in  $\bar{f}_\theta(t)$ and $- \epsilon \chi_\perp H_z h_y\cos\varphi_0$ in $\bar{f}_\varphi(t)$ because they are small. The expressions \eqref{eq:drivingforces} display a weak field-derivative Zeeman torque $\propto \epsilon \dot{h}_x$  (where $\epsilon \ll 1$) when the THz magnetic field is along the $x$-axis ($\dot h_y = 0, \dot h_x \neq 0$), being even zero when there is no tilt of the field $\epsilon = 0$. On the other hand, there is a strong field-derivative torque $\propto \dot{h}_y$ when the THz magnetic field is aligned along the $y$-axis, which is perpendicular to the equilibrium spin-direction and thus $\mathbf{M}$. In our previous article \cite{PhysRevLett.127.037203}, this precise torque was found to be responsible for the excitation of the exchange mode. Indeed, it was confirmed both experimentally and numerically that the tilt of the field is required for the excitation when $\mathbf{h}_{\mathrm{THz}} \parallel \hat{\mathbf{x}}$, while the excitation was optimal when $\mathbf{h}_{\mathrm{THz}} \perp \mathbf{M}$.
\end{widetext}


\clearpage

\bibliography{references}

\begin{thebibliography}{56}%
\makeatletter
\providecommand \@ifxundefined [1]{%
 \@ifx{#1\undefined}
}%
\providecommand \@ifnum [1]{%
 \ifnum #1\expandafter \@firstoftwo
 \else \expandafter \@secondoftwo
 \fi
}%
\providecommand \@ifx [1]{%
 \ifx #1\expandafter \@firstoftwo
 \else \expandafter \@secondoftwo
 \fi
}%
\providecommand \natexlab [1]{#1}%
\providecommand \enquote  [1]{``#1''}%
\providecommand \bibnamefont  [1]{#1}%
\providecommand \bibfnamefont [1]{#1}%
\providecommand \citenamefont [1]{#1}%
\providecommand \href@noop [0]{\@secondoftwo}%
\providecommand \href [0]{\begingroup \@sanitize@url \@href}%
\providecommand \@href[1]{\@@startlink{#1}\@@href}%
\providecommand \@@href[1]{\endgroup#1\@@endlink}%
\providecommand \@sanitize@url [0]{\catcode `\\12\catcode `\$12\catcode
  `\&12\catcode `\#12\catcode `\^12\catcode `\_12\catcode `\%12\relax}%
\providecommand \@@startlink[1]{}%
\providecommand \@@endlink[0]{}%
\providecommand \url  [0]{\begingroup\@sanitize@url \@url }%
\providecommand \@url [1]{\endgroup\@href {#1}{\urlprefix }}%
\providecommand \urlprefix  [0]{URL }%
\providecommand \Eprint [0]{\href }%
\providecommand \doibase [0]{https://doi.org/}%
\providecommand \selectlanguage [0]{\@gobble}%
\providecommand \bibinfo  [0]{\@secondoftwo}%
\providecommand \bibfield  [0]{\@secondoftwo}%
\providecommand \translation [1]{[#1]}%
\providecommand \BibitemOpen [0]{}%
\providecommand \bibitemStop [0]{}%
\providecommand \bibitemNoStop [0]{.\EOS\space}%
\providecommand \EOS [0]{\spacefactor3000\relax}%
\providecommand \BibitemShut  [1]{\csname bibitem#1\endcsname}%
\let\auto@bib@innerbib\@empty
\bibitem [{\citenamefont {Beaurepaire}\ \emph {et~al.}(1996)\citenamefont
  {Beaurepaire}, \citenamefont {Merle}, \citenamefont {Daunois},\ and\
  \citenamefont {Bigot}}]{PhysRevLett.76.4250}%
  \BibitemOpen
  \bibfield  {author} {\bibinfo {author} {\bibfnamefont {E.}~\bibnamefont
  {Beaurepaire}}, \bibinfo {author} {\bibfnamefont {J.-C.}\ \bibnamefont
  {Merle}}, \bibinfo {author} {\bibfnamefont {A.}~\bibnamefont {Daunois}},\
  and\ \bibinfo {author} {\bibfnamefont {J.-Y.}\ \bibnamefont {Bigot}},\
  }\bibfield  {title} {\bibinfo {title} {Ultrafast spin dynamics in
  ferromagnetic nickel},\ }\href {https://doi.org/10.1103/PhysRevLett.76.4250}
  {\bibfield  {journal} {\bibinfo  {journal} {Phys. Rev. Lett.}\ }\textbf
  {\bibinfo {volume} {76}},\ \bibinfo {pages} {4250} (\bibinfo {year}
  {1996})}\BibitemShut {NoStop}%
\bibitem [{\citenamefont {Curie}(1894)}]{curie1894symetrie}%
  \BibitemOpen
  \bibfield  {author} {\bibinfo {author} {\bibfnamefont {P.}~\bibnamefont
  {Curie}},\ }\bibfield  {title} {\bibinfo {title} {Sur la sym{\'e}trie dans
  les ph{\'e}nom{\`e}nes physiques, sym{\'e}trie d'un champ {\'e}lectrique et
  d'un champ magn{\'e}tique},\ }\href@noop {} {\bibfield  {journal} {\bibinfo
  {journal} {Journal de physique th{\'e}orique et appliqu{\'e}e}\ }\textbf
  {\bibinfo {volume} {3}},\ \bibinfo {pages} {393} (\bibinfo {year}
  {1894})}\BibitemShut {NoStop}%
\bibitem [{\citenamefont {Stanciu}\ \emph {et~al.}(2007)\citenamefont
  {Stanciu}, \citenamefont {Hansteen}, \citenamefont {Kimel}, \citenamefont
  {Kirilyuk}, \citenamefont {Tsukamoto}, \citenamefont {Itoh},\ and\
  \citenamefont {Rasing}}]{PhysRevLett.99.047601}%
  \BibitemOpen
  \bibfield  {author} {\bibinfo {author} {\bibfnamefont {C.~D.}\ \bibnamefont
  {Stanciu}}, \bibinfo {author} {\bibfnamefont {F.}~\bibnamefont {Hansteen}},
  \bibinfo {author} {\bibfnamefont {A.~V.}\ \bibnamefont {Kimel}}, \bibinfo
  {author} {\bibfnamefont {A.}~\bibnamefont {Kirilyuk}}, \bibinfo {author}
  {\bibfnamefont {A.}~\bibnamefont {Tsukamoto}}, \bibinfo {author}
  {\bibfnamefont {A.}~\bibnamefont {Itoh}},\ and\ \bibinfo {author}
  {\bibfnamefont {T.}~\bibnamefont {Rasing}},\ }\bibfield  {title} {\bibinfo
  {title} {All-optical magnetic recording with circularly polarized light},\
  }\href {https://doi.org/10.1103/PhysRevLett.99.047601} {\bibfield  {journal}
  {\bibinfo  {journal} {Phys. Rev. Lett.}\ }\textbf {\bibinfo {volume} {99}},\
  \bibinfo {pages} {047601} (\bibinfo {year} {2007})}\BibitemShut {NoStop}%
\bibitem [{\citenamefont {Radu}\ \emph {et~al.}(2011)\citenamefont {Radu},
  \citenamefont {Vahaplar}, \citenamefont {Stamm}, \citenamefont {Kachel},
  \citenamefont {Pontius}, \citenamefont {D{\"u}rr}, \citenamefont {Ostler},
  \citenamefont {Barker}, \citenamefont {Evans}, \citenamefont {Chantrell},
  \citenamefont {Tsukamoto}, \citenamefont {Itoh}, \citenamefont {Kirilyuk},
  \citenamefont {Rasing},\ and\ \citenamefont {Kimel}}]{Radu2011}%
  \BibitemOpen
  \bibfield  {author} {\bibinfo {author} {\bibfnamefont {I.}~\bibnamefont
  {Radu}}, \bibinfo {author} {\bibfnamefont {K.}~\bibnamefont {Vahaplar}},
  \bibinfo {author} {\bibfnamefont {C.}~\bibnamefont {Stamm}}, \bibinfo
  {author} {\bibfnamefont {T.}~\bibnamefont {Kachel}}, \bibinfo {author}
  {\bibfnamefont {N.}~\bibnamefont {Pontius}}, \bibinfo {author} {\bibfnamefont
  {H.~A.}\ \bibnamefont {D{\"u}rr}}, \bibinfo {author} {\bibfnamefont {T.~A.}\
  \bibnamefont {Ostler}}, \bibinfo {author} {\bibfnamefont {J.}~\bibnamefont
  {Barker}}, \bibinfo {author} {\bibfnamefont {R.~F.~L.}\ \bibnamefont
  {Evans}}, \bibinfo {author} {\bibfnamefont {R.~W.}\ \bibnamefont
  {Chantrell}}, \bibinfo {author} {\bibfnamefont {A.}~\bibnamefont
  {Tsukamoto}}, \bibinfo {author} {\bibfnamefont {A.}~\bibnamefont {Itoh}},
  \bibinfo {author} {\bibfnamefont {A.}~\bibnamefont {Kirilyuk}}, \bibinfo
  {author} {\bibfnamefont {T.}~\bibnamefont {Rasing}},\ and\ \bibinfo {author}
  {\bibfnamefont {A.~V.}\ \bibnamefont {Kimel}},\ }\bibfield  {title} {\bibinfo
  {title} {Transient ferromagnetic-like state mediating ultrafast reversal of
  antiferromagnetically coupled spins},\ }\href
  {https://doi.org/10.1038/nature09901} {\bibfield  {journal} {\bibinfo
  {journal} {Nature}\ }\textbf {\bibinfo {volume} {472}},\ \bibinfo {pages}
  {205} (\bibinfo {year} {2011})}\BibitemShut {NoStop}%
\bibitem [{\citenamefont {Ostler}\ \emph {et~al.}(2012)\citenamefont {Ostler},
  \citenamefont {Barker}, \citenamefont {Evans}, \citenamefont {Chantrell},
  \citenamefont {Atxitia}, \citenamefont {Chubykalo-Fesenko}, \citenamefont
  {El~Moussaoui}, \citenamefont {Le~Guyader}, \citenamefont {Mengotti},
  \citenamefont {Heyderman}, \citenamefont {Nolting}, \citenamefont
  {Tsukamoto}, \citenamefont {Itoh}, \citenamefont {Afanasiev}, \citenamefont
  {Ivanov}, \citenamefont {Kalashnikova}, \citenamefont {Vahaplar},
  \citenamefont {Mentink}, \citenamefont {Kirilyuk}, \citenamefont {Rasing},\
  and\ \citenamefont {Kimel}}]{Ostler2012}%
  \BibitemOpen
  \bibfield  {author} {\bibinfo {author} {\bibfnamefont {T.~A.}\ \bibnamefont
  {Ostler}}, \bibinfo {author} {\bibfnamefont {J.}~\bibnamefont {Barker}},
  \bibinfo {author} {\bibfnamefont {R.~F.~L.}\ \bibnamefont {Evans}}, \bibinfo
  {author} {\bibfnamefont {R.~W.}\ \bibnamefont {Chantrell}}, \bibinfo {author}
  {\bibfnamefont {U.}~\bibnamefont {Atxitia}}, \bibinfo {author} {\bibfnamefont
  {O.}~\bibnamefont {Chubykalo-Fesenko}}, \bibinfo {author} {\bibfnamefont
  {S.}~\bibnamefont {El~Moussaoui}}, \bibinfo {author} {\bibfnamefont
  {L.}~\bibnamefont {Le~Guyader}}, \bibinfo {author} {\bibfnamefont
  {E.}~\bibnamefont {Mengotti}}, \bibinfo {author} {\bibfnamefont {L.~J.}\
  \bibnamefont {Heyderman}}, \bibinfo {author} {\bibfnamefont {F.}~\bibnamefont
  {Nolting}}, \bibinfo {author} {\bibfnamefont {A.}~\bibnamefont {Tsukamoto}},
  \bibinfo {author} {\bibfnamefont {A.}~\bibnamefont {Itoh}}, \bibinfo {author}
  {\bibfnamefont {D.}~\bibnamefont {Afanasiev}}, \bibinfo {author}
  {\bibfnamefont {B.~A.}\ \bibnamefont {Ivanov}}, \bibinfo {author}
  {\bibfnamefont {A.~M.}\ \bibnamefont {Kalashnikova}}, \bibinfo {author}
  {\bibfnamefont {K.}~\bibnamefont {Vahaplar}}, \bibinfo {author}
  {\bibfnamefont {J.}~\bibnamefont {Mentink}}, \bibinfo {author} {\bibfnamefont
  {A.}~\bibnamefont {Kirilyuk}}, \bibinfo {author} {\bibfnamefont
  {T.}~\bibnamefont {Rasing}},\ and\ \bibinfo {author} {\bibfnamefont {A.~V.}\
  \bibnamefont {Kimel}},\ }\bibfield  {title} {\bibinfo {title} {Ultrafast
  heating as a sufficient stimulus for magnetization reversal in a
  ferrimagnet},\ }\href {https://doi.org/10.1038/ncomms1666} {\bibfield
  {journal} {\bibinfo  {journal} {Nature Communications}\ }\textbf {\bibinfo
  {volume} {3}},\ \bibinfo {pages} {666} (\bibinfo {year} {2012})}\BibitemShut
  {NoStop}%
\bibitem [{\citenamefont {Yang}\ \emph {et~al.}(2017)\citenamefont {Yang},
  \citenamefont {Wilson}, \citenamefont {Gorchon}, \citenamefont {Lambert},
  \citenamefont {Salahuddin},\ and\ \citenamefont
  {Bokor}}]{doi:10.1126/sciadv.1603117}%
  \BibitemOpen
  \bibfield  {author} {\bibinfo {author} {\bibfnamefont {Y.}~\bibnamefont
  {Yang}}, \bibinfo {author} {\bibfnamefont {R.~B.}\ \bibnamefont {Wilson}},
  \bibinfo {author} {\bibfnamefont {J.}~\bibnamefont {Gorchon}}, \bibinfo
  {author} {\bibfnamefont {C.-H.}\ \bibnamefont {Lambert}}, \bibinfo {author}
  {\bibfnamefont {S.}~\bibnamefont {Salahuddin}},\ and\ \bibinfo {author}
  {\bibfnamefont {J.}~\bibnamefont {Bokor}},\ }\bibfield  {title} {\bibinfo
  {title} {Ultrafast magnetization reversal by picosecond electrical pulses},\
  }\href {https://doi.org/10.1126/sciadv.1603117} {\bibfield  {journal}
  {\bibinfo  {journal} {Sci. Adv.}\ }\textbf {\bibinfo {volume} {3}},\ \bibinfo
  {pages} {e1603117} (\bibinfo {year} {2017})}\BibitemShut {NoStop}%
\bibitem [{\citenamefont {Stupakiewicz}\ \emph {et~al.}(2017)\citenamefont
  {Stupakiewicz}, \citenamefont {Szerenos}, \citenamefont {Afanasiev},
  \citenamefont {Kirilyuk},\ and\ \citenamefont {Kimel}}]{Stupakiewicz2017}%
  \BibitemOpen
  \bibfield  {author} {\bibinfo {author} {\bibfnamefont {A.}~\bibnamefont
  {Stupakiewicz}}, \bibinfo {author} {\bibfnamefont {K.}~\bibnamefont
  {Szerenos}}, \bibinfo {author} {\bibfnamefont {D.}~\bibnamefont {Afanasiev}},
  \bibinfo {author} {\bibfnamefont {A.}~\bibnamefont {Kirilyuk}},\ and\
  \bibinfo {author} {\bibfnamefont {A.~V.}\ \bibnamefont {Kimel}},\ }\bibfield
  {title} {\bibinfo {title} {Ultrafast nonthermal photo-magnetic recording in a
  transparent medium},\ }\href {https://doi.org/10.1038/nature20807} {\bibfield
   {journal} {\bibinfo  {journal} {Nature}\ }\textbf {\bibinfo {volume}
  {542}},\ \bibinfo {pages} {71} (\bibinfo {year} {2017})}\BibitemShut
  {NoStop}%
\bibitem [{\citenamefont {Hebling}\ \emph {et~al.}(2002)\citenamefont
  {Hebling}, \citenamefont {Alm\'{a}si}, \citenamefont {Kozma},\ and\
  \citenamefont {Kuhl}}]{Hebling:02}%
  \BibitemOpen
  \bibfield  {author} {\bibinfo {author} {\bibfnamefont {J.}~\bibnamefont
  {Hebling}}, \bibinfo {author} {\bibfnamefont {G.}~\bibnamefont {Alm\'{a}si}},
  \bibinfo {author} {\bibfnamefont {I.~Z.}\ \bibnamefont {Kozma}},\ and\
  \bibinfo {author} {\bibfnamefont {J.}~\bibnamefont {Kuhl}},\ }\bibfield
  {title} {\bibinfo {title} {Velocity matching by pulse front tilting for
  large-area {THz}-pulse generation},\ }\href
  {https://doi.org/10.1364/OE.10.001161} {\bibfield  {journal} {\bibinfo
  {journal} {Opt. Express}\ }\textbf {\bibinfo {volume} {10}},\ \bibinfo
  {pages} {1161} (\bibinfo {year} {2002})}\BibitemShut {NoStop}%
\bibitem [{\citenamefont {Kampfrath}\ \emph {et~al.}(2011)\citenamefont
  {Kampfrath}, \citenamefont {Sell}, \citenamefont {Klatt}, \citenamefont
  {Pashkin}, \citenamefont {M{\"a}hrlein}, \citenamefont {Dekorsy},
  \citenamefont {Wolf}, \citenamefont {Fiebig}, \citenamefont {Leitenstorfer},\
  and\ \citenamefont {Huber}}]{Kampfrath2011}%
  \BibitemOpen
  \bibfield  {author} {\bibinfo {author} {\bibfnamefont {T.}~\bibnamefont
  {Kampfrath}}, \bibinfo {author} {\bibfnamefont {A.}~\bibnamefont {Sell}},
  \bibinfo {author} {\bibfnamefont {G.}~\bibnamefont {Klatt}}, \bibinfo
  {author} {\bibfnamefont {A.}~\bibnamefont {Pashkin}}, \bibinfo {author}
  {\bibfnamefont {S.}~\bibnamefont {M{\"a}hrlein}}, \bibinfo {author}
  {\bibfnamefont {T.}~\bibnamefont {Dekorsy}}, \bibinfo {author} {\bibfnamefont
  {M.}~\bibnamefont {Wolf}}, \bibinfo {author} {\bibfnamefont {M.}~\bibnamefont
  {Fiebig}}, \bibinfo {author} {\bibfnamefont {A.}~\bibnamefont
  {Leitenstorfer}},\ and\ \bibinfo {author} {\bibfnamefont {R.}~\bibnamefont
  {Huber}},\ }\bibfield  {title} {\bibinfo {title} {Coherent terahertz control
  of antiferromagnetic spin waves},\ }\href
  {https://doi.org/10.1038/nphoton.2010.259} {\bibfield  {journal} {\bibinfo
  {journal} {Nat. Photonics}\ }\textbf {\bibinfo {volume} {5}},\ \bibinfo
  {pages} {31} (\bibinfo {year} {2011})}\BibitemShut {NoStop}%
\bibitem [{\citenamefont {Baierl}\ \emph {et~al.}(2016)\citenamefont {Baierl},
  \citenamefont {Hohenleutner}, \citenamefont {Kampfrath} \emph
  {et~al.}}]{nonspincontrol}%
  \BibitemOpen
  \bibfield  {author} {\bibinfo {author} {\bibfnamefont {S.}~\bibnamefont
  {Baierl}}, \bibinfo {author} {\bibfnamefont {M.}~\bibnamefont
  {Hohenleutner}}, \bibinfo {author} {\bibfnamefont {T.}~\bibnamefont
  {Kampfrath}}, \emph {et~al.},\ }\bibfield  {title} {\bibinfo {title}
  {Nonlinear spin control by terahertz-driven anisotropy fields},\ }\href@noop
  {} {\bibfield  {journal} {\bibinfo  {journal} {Nat. Photonics}\ }\textbf
  {\bibinfo {volume} {10}},\ \bibinfo {pages} {715} (\bibinfo {year}
  {2016})}\BibitemShut {NoStop}%
\bibitem [{\citenamefont {Schlauderer}\ \emph {et~al.}(2019)\citenamefont
  {Schlauderer}, \citenamefont {Lange}, \citenamefont {Baierl}, \citenamefont
  {Ebnet}, \citenamefont {Schmid}, \citenamefont {Valovcin}, \citenamefont
  {Zvezdin}, \citenamefont {Kimel}, \citenamefont {Mikhaylovskiy},\ and\
  \citenamefont {Huber}}]{Schlauderer2019}%
  \BibitemOpen
  \bibfield  {author} {\bibinfo {author} {\bibfnamefont {S.}~\bibnamefont
  {Schlauderer}}, \bibinfo {author} {\bibfnamefont {C.}~\bibnamefont {Lange}},
  \bibinfo {author} {\bibfnamefont {S.}~\bibnamefont {Baierl}}, \bibinfo
  {author} {\bibfnamefont {T.}~\bibnamefont {Ebnet}}, \bibinfo {author}
  {\bibfnamefont {C.~P.}\ \bibnamefont {Schmid}}, \bibinfo {author}
  {\bibfnamefont {D.~C.}\ \bibnamefont {Valovcin}}, \bibinfo {author}
  {\bibfnamefont {A.~K.}\ \bibnamefont {Zvezdin}}, \bibinfo {author}
  {\bibfnamefont {A.~V.}\ \bibnamefont {Kimel}}, \bibinfo {author}
  {\bibfnamefont {R.~V.}\ \bibnamefont {Mikhaylovskiy}},\ and\ \bibinfo
  {author} {\bibfnamefont {R.}~\bibnamefont {Huber}},\ }\bibfield  {title}
  {\bibinfo {title} {Temporal and spectral fingerprints of ultrafast
  all-coherent spin switching},\ }\href
  {https://doi.org/10.1038/s41586-019-1174-7} {\bibfield  {journal} {\bibinfo
  {journal} {Nature}\ }\textbf {\bibinfo {volume} {569}},\ \bibinfo {pages}
  {383} (\bibinfo {year} {2019})}\BibitemShut {NoStop}%
\bibitem [{\citenamefont {Mashkovich}\ \emph {et~al.}(2019)\citenamefont
  {Mashkovich}, \citenamefont {Grishunin}, \citenamefont {Mikhaylovskiy},
  \citenamefont {Zvezdin}, \citenamefont {Pisarev}, \citenamefont {Strugatsky},
  \citenamefont {Christianen}, \citenamefont {Rasing},\ and\ \citenamefont
  {Kimel}}]{PhysRevLett.123.157202}%
  \BibitemOpen
  \bibfield  {author} {\bibinfo {author} {\bibfnamefont {E.~A.}\ \bibnamefont
  {Mashkovich}}, \bibinfo {author} {\bibfnamefont {K.~A.}\ \bibnamefont
  {Grishunin}}, \bibinfo {author} {\bibfnamefont {R.~V.}\ \bibnamefont
  {Mikhaylovskiy}}, \bibinfo {author} {\bibfnamefont {A.~K.}\ \bibnamefont
  {Zvezdin}}, \bibinfo {author} {\bibfnamefont {R.~V.}\ \bibnamefont
  {Pisarev}}, \bibinfo {author} {\bibfnamefont {M.~B.}\ \bibnamefont
  {Strugatsky}}, \bibinfo {author} {\bibfnamefont {P.~C.~M.}\ \bibnamefont
  {Christianen}}, \bibinfo {author} {\bibfnamefont {T.}~\bibnamefont
  {Rasing}},\ and\ \bibinfo {author} {\bibfnamefont {A.~V.}\ \bibnamefont
  {Kimel}},\ }\bibfield  {title} {\bibinfo {title} {Terahertz optomagnetism:
  Nonlinear {THz} excitation of {GHz} spin waves in antiferromagnetic
  {${\mathrm{FeBO}}_{3}$}},\ }\href
  {https://doi.org/10.1103/PhysRevLett.123.157202} {\bibfield  {journal}
  {\bibinfo  {journal} {Phys. Rev. Lett.}\ }\textbf {\bibinfo {volume} {123}},\
  \bibinfo {pages} {157202} (\bibinfo {year} {2019})}\BibitemShut {NoStop}%
\bibitem [{\citenamefont {Blank}\ \emph {et~al.}(2021)\citenamefont {Blank},
  \citenamefont {Grishunin}, \citenamefont {Mashkovich}, \citenamefont
  {Logunov}, \citenamefont {Zvezdin},\ and\ \citenamefont
  {Kimel}}]{PhysRevLett.127.037203}%
  \BibitemOpen
  \bibfield  {author} {\bibinfo {author} {\bibfnamefont {T.~G.~H.}\
  \bibnamefont {Blank}}, \bibinfo {author} {\bibfnamefont {K.~A.}\ \bibnamefont
  {Grishunin}}, \bibinfo {author} {\bibfnamefont {E.~A.}\ \bibnamefont
  {Mashkovich}}, \bibinfo {author} {\bibfnamefont {M.~V.}\ \bibnamefont
  {Logunov}}, \bibinfo {author} {\bibfnamefont {A.~K.}\ \bibnamefont
  {Zvezdin}},\ and\ \bibinfo {author} {\bibfnamefont {A.~V.}\ \bibnamefont
  {Kimel}},\ }\bibfield  {title} {\bibinfo {title} {Thz-scale field-induced
  spin dynamics in ferrimagnetic iron garnets},\ }\href
  {https://doi.org/10.1103/PhysRevLett.127.037203} {\bibfield  {journal}
  {\bibinfo  {journal} {Phys. Rev. Lett.}\ }\textbf {\bibinfo {volume} {127}},\
  \bibinfo {pages} {037203} (\bibinfo {year} {2021})}\BibitemShut {NoStop}%
\bibitem [{\citenamefont {Zon}\ \emph {et~al.}(1987)\citenamefont {Zon},
  \citenamefont {Kupershmidt}, \citenamefont {Pakhomov},\ and\ \citenamefont
  {Urazbaev}}]{ICMEobersvation}%
  \BibitemOpen
  \bibfield  {author} {\bibinfo {author} {\bibfnamefont {B.~A.}\ \bibnamefont
  {Zon}}, \bibinfo {author} {\bibfnamefont {V.~Y.}\ \bibnamefont
  {Kupershmidt}}, \bibinfo {author} {\bibfnamefont {G.~V.}\ \bibnamefont
  {Pakhomov}},\ and\ \bibinfo {author} {\bibfnamefont {T.~T.}\ \bibnamefont
  {Urazbaev}},\ }\bibfield  {title} {\bibinfo {title} {Observation of inverse
  {Cotton-Mouton} effect in the magnetically ordered crystal
  {(Lu,B)\textsubscript{3}(Fe, Ga)\textsubscript{5}O\textsubscript{12}}},\
  }\href@noop {} {\bibfield  {journal} {\bibinfo  {journal} {JETP Lett.}\
  }\textbf {\bibinfo {volume} {45}},\ \bibinfo {pages} {272} (\bibinfo {year}
  {1987})}\BibitemShut {NoStop}%
\bibitem [{\citenamefont {Shen}\ \emph {et~al.}(2018)\citenamefont {Shen},
  \citenamefont {Zhou}, \citenamefont {Shi}, \citenamefont {Tang},
  \citenamefont {Zheng}, \citenamefont {Wu}, \citenamefont {Zhou},
  \citenamefont {Chen},\ and\ \citenamefont {Zhao}}]{PhysRevB.97.224430}%
  \BibitemOpen
  \bibfield  {author} {\bibinfo {author} {\bibfnamefont {L.~Q.}\ \bibnamefont
  {Shen}}, \bibinfo {author} {\bibfnamefont {L.~F.}\ \bibnamefont {Zhou}},
  \bibinfo {author} {\bibfnamefont {J.~Y.}\ \bibnamefont {Shi}}, \bibinfo
  {author} {\bibfnamefont {M.}~\bibnamefont {Tang}}, \bibinfo {author}
  {\bibfnamefont {Z.}~\bibnamefont {Zheng}}, \bibinfo {author} {\bibfnamefont
  {D.}~\bibnamefont {Wu}}, \bibinfo {author} {\bibfnamefont {S.~M.}\
  \bibnamefont {Zhou}}, \bibinfo {author} {\bibfnamefont {L.~Y.}\ \bibnamefont
  {Chen}},\ and\ \bibinfo {author} {\bibfnamefont {H.~B.}\ \bibnamefont
  {Zhao}},\ }\bibfield  {title} {\bibinfo {title} {Dominant role of inverse
  {Cotton-Mouton} effect in ultrafast stimulation of magnetization precession
  in undoped yttrium iron garnet films by 400-nm laser pulses},\ }\href
  {https://doi.org/10.1103/PhysRevB.97.224430} {\bibfield  {journal} {\bibinfo
  {journal} {Phys. Rev. B}\ }\textbf {\bibinfo {volume} {97}},\ \bibinfo
  {pages} {224430} (\bibinfo {year} {2018})}\BibitemShut {NoStop}%
\bibitem [{\citenamefont {Davydova}\ \emph {et~al.}(2019)\citenamefont
  {Davydova}, \citenamefont {Zvezdin}, \citenamefont {Kimel},\ and\
  \citenamefont {Zvezdin}}]{Davydova_2019}%
  \BibitemOpen
  \bibfield  {author} {\bibinfo {author} {\bibfnamefont {M.~D.}\ \bibnamefont
  {Davydova}}, \bibinfo {author} {\bibfnamefont {K.~A.}\ \bibnamefont
  {Zvezdin}}, \bibinfo {author} {\bibfnamefont {A.~V.}\ \bibnamefont {Kimel}},\
  and\ \bibinfo {author} {\bibfnamefont {A.~K.}\ \bibnamefont {Zvezdin}},\
  }\bibfield  {title} {\bibinfo {title} {Ultrafast spin dynamics in
  ferrimagnets with compensation point},\ }\href
  {https://doi.org/10.1088/1361-648x/ab42fa} {\bibfield  {journal} {\bibinfo
  {journal} {J. Phys.: Condensed Matter}\ }\textbf {\bibinfo {volume} {32}},\
  \bibinfo {pages} {01LT01} (\bibinfo {year} {2019})}\BibitemShut {NoStop}%
\bibitem [{\citenamefont {Shen}\ and\ \citenamefont
  {Bloembergen}(1966)}]{PhysRev.143.372}%
  \BibitemOpen
  \bibfield  {author} {\bibinfo {author} {\bibfnamefont {Y.~R.}\ \bibnamefont
  {Shen}}\ and\ \bibinfo {author} {\bibfnamefont {N.}~\bibnamefont
  {Bloembergen}},\ }\bibfield  {title} {\bibinfo {title} {Interaction between
  light waves and spin waves},\ }\href
  {https://doi.org/10.1103/PhysRev.143.372} {\bibfield  {journal} {\bibinfo
  {journal} {Phys. Rev.}\ }\textbf {\bibinfo {volume} {143}},\ \bibinfo {pages}
  {372} (\bibinfo {year} {1966})}\BibitemShut {NoStop}%
\bibitem [{\citenamefont {Lu}\ \emph {et~al.}(2018)\citenamefont {Lu},
  \citenamefont {Li}, \citenamefont {Zhang}, \citenamefont {Hwang},
  \citenamefont {Ofori-Okai},\ and\ \citenamefont {Nelson}}]{Lu2018}%
  \BibitemOpen
  \bibfield  {author} {\bibinfo {author} {\bibfnamefont {J.}~\bibnamefont
  {Lu}}, \bibinfo {author} {\bibfnamefont {X.}~\bibnamefont {Li}}, \bibinfo
  {author} {\bibfnamefont {Y.}~\bibnamefont {Zhang}}, \bibinfo {author}
  {\bibfnamefont {H.~Y.}\ \bibnamefont {Hwang}}, \bibinfo {author}
  {\bibfnamefont {B.~K.}\ \bibnamefont {Ofori-Okai}},\ and\ \bibinfo {author}
  {\bibfnamefont {K.~A.}\ \bibnamefont {Nelson}},\ }\bibfield  {title}
  {\bibinfo {title} {Two-dimensional spectroscopy at terahertz frequencies},\
  }\href {https://doi.org/10.1007/s41061-018-0185-4} {\bibfield  {journal}
  {\bibinfo  {journal} {Top Curr Chem (Z)}\ }\textbf {\bibinfo {volume}
  {376}},\ \bibinfo {pages} {6} (\bibinfo {year} {2018})}\BibitemShut {NoStop}%
\bibitem [{\citenamefont {Geller}\ and\ \citenamefont
  {Gilleo}(1957)}]{GELLER195730}%
  \BibitemOpen
  \bibfield  {author} {\bibinfo {author} {\bibfnamefont {S.}~\bibnamefont
  {Geller}}\ and\ \bibinfo {author} {\bibfnamefont {M.}~\bibnamefont
  {Gilleo}},\ }\bibfield  {title} {\bibinfo {title} {The crystal structure and
  ferrimagnetism of yttrium-iron garnet, {Y3Fe2(FeO4)3}},\ }\href
  {https://doi.org/https://doi.org/10.1016/0022-3697(57)90044-6} {\bibfield
  {journal} {\bibinfo  {journal} {Journal of Physics and Chemistry of Solids}\
  }\textbf {\bibinfo {volume} {3}},\ \bibinfo {pages} {30} (\bibinfo {year}
  {1957})}\BibitemShut {NoStop}%
\bibitem [{\citenamefont {Wohlfarth}(1986)}]{wohlfarth1986handbook}%
  \BibitemOpen
  \bibfield  {author} {\bibinfo {author} {\bibfnamefont {E.~P.}\ \bibnamefont
  {Wohlfarth}},\ }\href@noop {} {\emph {\bibinfo {title} {Handbook of magnetic
  materials}}},\ Vol.~\bibinfo {volume} {2}\ (\bibinfo  {publisher}
  {Elsevier},\ \bibinfo {year} {1986})\ Chap.~\bibinfo {chapter}
  {2}\BibitemShut {NoStop}%
\bibitem [{\citenamefont {Levitin}\ \emph {et~al.}(1971)\citenamefont
  {Levitin}, \citenamefont {Ponomarev},\ and\ \citenamefont {Popov}}]{Levitin}%
  \BibitemOpen
  \bibfield  {author} {\bibinfo {author} {\bibfnamefont {R.}~\bibnamefont
  {Levitin}}, \bibinfo {author} {\bibfnamefont {B.}~\bibnamefont {Ponomarev}},\
  and\ \bibinfo {author} {\bibfnamefont {Y.}~\bibnamefont {Popov}},\ }\bibfield
   {title} {\bibinfo {title} {Magnetization of iron garnets of heavy rare earth
  elements in fields up to 240 {kOe}},\ }\href@noop {} {\bibfield  {journal}
  {\bibinfo  {journal} {JETP}\ }\textbf {\bibinfo {volume} {32}},\ \bibinfo
  {pages} {1056} (\bibinfo {year} {1971})}\BibitemShut {NoStop}%
\bibitem [{\citenamefont {Clark}\ and\ \citenamefont
  {Callen}(1968)}]{neelferri}%
  \BibitemOpen
  \bibfield  {author} {\bibinfo {author} {\bibfnamefont {A.~E.}\ \bibnamefont
  {Clark}}\ and\ \bibinfo {author} {\bibfnamefont {E.}~\bibnamefont {Callen}},\
  }\bibfield  {title} {\bibinfo {title} {Néel ferrimagnets in large magnetic
  fields},\ }\href {https://doi.org/10.1063/1.1656100} {\bibfield  {journal}
  {\bibinfo  {journal} {Journal of Applied Physics}\ }\textbf {\bibinfo
  {volume} {39}},\ \bibinfo {pages} {5972} (\bibinfo {year}
  {1968})}\BibitemShut {NoStop}%
\bibitem [{\citenamefont {Hibiya}\ \emph {et~al.}(1985)\citenamefont {Hibiya},
  \citenamefont {Morishige},\ and\ \citenamefont {Nakashima}}]{Hibiya_1985}%
  \BibitemOpen
  \bibfield  {author} {\bibinfo {author} {\bibfnamefont {T.}~\bibnamefont
  {Hibiya}}, \bibinfo {author} {\bibfnamefont {Y.}~\bibnamefont {Morishige}},\
  and\ \bibinfo {author} {\bibfnamefont {J.}~\bibnamefont {Nakashima}},\
  }\bibfield  {title} {\bibinfo {title} {Growth and characterization of
  liquid-phase epitaxial bi-substituted iron garnet films for magneto-optic
  application},\ }\href {https://doi.org/10.1143/JJAP.24.1316} {\bibfield
  {journal} {\bibinfo  {journal} {Japanese Journal of Applied Physics}\
  }\textbf {\bibinfo {volume} {24}},\ \bibinfo {pages} {1316} (\bibinfo {year}
  {1985})}\BibitemShut {NoStop}%
\bibitem [{\citenamefont {Hansen}\ and\ \citenamefont
  {Tolksdorf}(1991)}]{Hansen}%
  \BibitemOpen
  \bibfield  {author} {\bibinfo {author} {\bibfnamefont {P.}~\bibnamefont
  {Hansen}}\ and\ \bibinfo {author} {\bibfnamefont {W.}~\bibnamefont
  {Tolksdorf}},\ }\bibfield  {title} {\bibinfo {title} {Magnetic and
  magneto‐optic properties of bismuth‐substituted thulium iron‐garnet
  films},\ }\href {https://doi.org/10.1063/1.348315} {\bibfield  {journal}
  {\bibinfo  {journal} {Journal of Applied Physics}\ }\textbf {\bibinfo
  {volume} {69}},\ \bibinfo {pages} {4577} (\bibinfo {year}
  {1991})}\BibitemShut {NoStop}%
\bibitem [{\citenamefont {Gerhardt}\ \emph {et~al.}(1993)\citenamefont
  {Gerhardt}, \citenamefont {Sure}, \citenamefont {Dötsch}, \citenamefont
  {Linkewitz},\ and\ \citenamefont
  {Tolksdorf}}]{TmBiFeGaO12_substitutionsgerhard}%
  \BibitemOpen
  \bibfield  {author} {\bibinfo {author} {\bibfnamefont {R.}~\bibnamefont
  {Gerhardt}}, \bibinfo {author} {\bibfnamefont {S.}~\bibnamefont {Sure}},
  \bibinfo {author} {\bibfnamefont {H.}~\bibnamefont {Dötsch}}, \bibinfo
  {author} {\bibfnamefont {T.}~\bibnamefont {Linkewitz}},\ and\ \bibinfo
  {author} {\bibfnamefont {W.}~\bibnamefont {Tolksdorf}},\ }\bibfield  {title}
  {\bibinfo {title} {Optical properties of bismuth and gallium substituted
  thulium iron garnet films},\ }\href
  {https://doi.org/https://doi.org/10.1016/0030-4018(93)90467-J} {\bibfield
  {journal} {\bibinfo  {journal} {Optics Communications}\ }\textbf {\bibinfo
  {volume} {102}},\ \bibinfo {pages} {31 } (\bibinfo {year}
  {1993})}\BibitemShut {NoStop}%
\bibitem [{\citenamefont {Geller}\ \emph {et~al.}(1966)\citenamefont {Geller},
  \citenamefont {Cape}, \citenamefont {Espinosa},\ and\ \citenamefont
  {Leslie}}]{PhysRev.148.522}%
  \BibitemOpen
  \bibfield  {author} {\bibinfo {author} {\bibfnamefont {S.}~\bibnamefont
  {Geller}}, \bibinfo {author} {\bibfnamefont {J.~A.}\ \bibnamefont {Cape}},
  \bibinfo {author} {\bibfnamefont {G.~P.}\ \bibnamefont {Espinosa}},\ and\
  \bibinfo {author} {\bibfnamefont {D.~H.}\ \bibnamefont {Leslie}},\ }\bibfield
   {title} {\bibinfo {title} {Gallium-substituted yttrium iron garnet},\ }\href
  {https://doi.org/10.1103/PhysRev.148.522} {\bibfield  {journal} {\bibinfo
  {journal} {Phys. Rev.}\ }\textbf {\bibinfo {volume} {148}},\ \bibinfo {pages}
  {522} (\bibinfo {year} {1966})}\BibitemShut {NoStop}%
\bibitem [{\citenamefont {Hirori}\ and\ \citenamefont
  {Tanaka}(2016)}]{Hirori_review_4ftheta}%
  \BibitemOpen
  \bibfield  {author} {\bibinfo {author} {\bibfnamefont {H.}~\bibnamefont
  {Hirori}}\ and\ \bibinfo {author} {\bibfnamefont {K.}~\bibnamefont
  {Tanaka}},\ }\bibfield  {title} {\bibinfo {title} {Dynamical nonlinear
  interactions of solids with strong terahertz pulses},\ }\href
  {https://doi.org/10.7566/JPSJ.85.082001} {\bibfield  {journal} {\bibinfo
  {journal} {Journal of the Physical Society of Japan}\ }\textbf {\bibinfo
  {volume} {85}},\ \bibinfo {pages} {082001} (\bibinfo {year}
  {2016})}\BibitemShut {NoStop}%
\bibitem [{Note1()}]{Note1}%
  \BibitemOpen
  \bibinfo {note} {\label {footnote}See the Supplemental Material for the XRD
  analysis of the sample, which includes Refs.~\cite {doi:10.1063/1.4929691,
  Ciubotariu2019}.}\BibitemShut {Stop}%
\bibitem [{\citenamefont {Kaplan}\ and\ \citenamefont
  {Kittel}(1953)}]{doi:10.1063/1.1699018}%
  \BibitemOpen
  \bibfield  {author} {\bibinfo {author} {\bibfnamefont {J.}~\bibnamefont
  {Kaplan}}\ and\ \bibinfo {author} {\bibfnamefont {C.}~\bibnamefont
  {Kittel}},\ }\bibfield  {title} {\bibinfo {title} {Exchange frequency
  electron spin resonance in ferrites},\ }\href@noop {} {\bibfield  {journal}
  {\bibinfo  {journal} {The Journal of Chemical Physics}\ }\textbf {\bibinfo
  {volume} {21}},\ \bibinfo {pages} {760} (\bibinfo {year} {1953})}\BibitemShut
  {NoStop}%
\bibitem [{\citenamefont {Gurevich}\ and\ \citenamefont
  {Melkov}(1996)}]{gurevich1996magnetization}%
  \BibitemOpen
  \bibfield  {author} {\bibinfo {author} {\bibfnamefont {A.}~\bibnamefont
  {Gurevich}}\ and\ \bibinfo {author} {\bibfnamefont {G.}~\bibnamefont
  {Melkov}},\ }\href {https://books.google.nl/books?id=YgQtSvFIvFQC} {\emph
  {\bibinfo {title} {Magnetization Oscillations and Waves}}}\ (\bibinfo
  {publisher} {Taylor \& Francis},\ \bibinfo {year} {1996})\BibitemShut
  {NoStop}%
\bibitem [{\citenamefont {Kittel}(1959)}]{PhysRev.115.1587}%
  \BibitemOpen
  \bibfield  {author} {\bibinfo {author} {\bibfnamefont {C.}~\bibnamefont
  {Kittel}},\ }\bibfield  {title} {\bibinfo {title} {Theory of ferromagnetic
  resonance in rare earth garnets. i. $g$ values},\ }\href
  {https://doi.org/10.1103/PhysRev.115.1587} {\bibfield  {journal} {\bibinfo
  {journal} {Phys. Rev.}\ }\textbf {\bibinfo {volume} {115}},\ \bibinfo {pages}
  {1587} (\bibinfo {year} {1959})}\BibitemShut {NoStop}%
\bibitem [{\citenamefont {Ohta}\ \emph {et~al.}(1977)\citenamefont {Ohta},
  \citenamefont {Ikeda}, \citenamefont {Ishida},\ and\ \citenamefont
  {Sugita}}]{doi:10.1143/JPSJ.43.705}%
  \BibitemOpen
  \bibfield  {author} {\bibinfo {author} {\bibfnamefont {N.}~\bibnamefont
  {Ohta}}, \bibinfo {author} {\bibfnamefont {T.}~\bibnamefont {Ikeda}},
  \bibinfo {author} {\bibfnamefont {F.}~\bibnamefont {Ishida}},\ and\ \bibinfo
  {author} {\bibfnamefont {Y.}~\bibnamefont {Sugita}},\ }\bibfield  {title}
  {\bibinfo {title} {High g bubble garnets without containing {Eu$^{3+}$}
  ion},\ }\href {https://doi.org/10.1143/JPSJ.43.705} {\bibfield  {journal}
  {\bibinfo  {journal} {Journal of the Physical Society of Japan}\ }\textbf
  {\bibinfo {volume} {43}},\ \bibinfo {pages} {705} (\bibinfo {year}
  {1977})}\BibitemShut {NoStop}%
\bibitem [{\citenamefont {Doroshev}\ \emph {et~al.}(1994)\citenamefont
  {Doroshev}, \citenamefont {Savosta},\ and\ \citenamefont
  {Novák}}]{DOROSHEV1994290}%
  \BibitemOpen
  \bibfield  {author} {\bibinfo {author} {\bibfnamefont {V.}~\bibnamefont
  {Doroshev}}, \bibinfo {author} {\bibfnamefont {M.}~\bibnamefont {Savosta}},\
  and\ \bibinfo {author} {\bibfnamefont {P.}~\bibnamefont {Novák}},\
  }\bibfield  {title} {\bibinfo {title} {On the magnetic structure of
  {(TmY)IG}},\ }\href
  {https://doi.org/https://doi.org/10.1016/0921-4526(94)90015-9} {\bibfield
  {journal} {\bibinfo  {journal} {Physica B: Condensed Matter}\ }\textbf
  {\bibinfo {volume} {198}},\ \bibinfo {pages} {290} (\bibinfo {year}
  {1994})}\BibitemShut {NoStop}%
\bibitem [{\citenamefont {Tcheou}\ \emph {et~al.}(1970)\citenamefont {Tcheou},
  \citenamefont {Bertaut},\ and\ \citenamefont {Fuess}}]{TCHEOU19701751}%
  \BibitemOpen
  \bibfield  {author} {\bibinfo {author} {\bibfnamefont {F.}~\bibnamefont
  {Tcheou}}, \bibinfo {author} {\bibfnamefont {E.}~\bibnamefont {Bertaut}},\
  and\ \bibinfo {author} {\bibfnamefont {H.}~\bibnamefont {Fuess}},\ }\bibfield
   {title} {\bibinfo {title} {Ii — neutron diffraction study of some rare
  earth iron garnets {RIG (R = Dy, Er, Yb, Tm)} at low temperatures},\ }\href
  {https://doi.org/https://doi.org/10.1016/0038-1098(70)90389-3} {\bibfield
  {journal} {\bibinfo  {journal} {Solid State Communications}\ }\textbf
  {\bibinfo {volume} {8}},\ \bibinfo {pages} {1751} (\bibinfo {year}
  {1970})}\BibitemShut {NoStop}%
\bibitem [{\citenamefont {Ciubotariu}\ \emph {et~al.}(2019)\citenamefont
  {Ciubotariu}, \citenamefont {Semisalova}, \citenamefont {Lenz},\ and\
  \citenamefont {Albrecht}}]{Ciubotariu2019}%
  \BibitemOpen
  \bibfield  {author} {\bibinfo {author} {\bibfnamefont {O.}~\bibnamefont
  {Ciubotariu}}, \bibinfo {author} {\bibfnamefont {A.}~\bibnamefont
  {Semisalova}}, \bibinfo {author} {\bibfnamefont {K.}~\bibnamefont {Lenz}},\
  and\ \bibinfo {author} {\bibfnamefont {M.}~\bibnamefont {Albrecht}},\
  }\bibfield  {title} {\bibinfo {title} {Strain-induced perpendicular magnetic
  anisotropy and gilbert damping of {Tm$_3$Fe$_5$O$_{12}$} thin films},\ }\href
  {https://doi.org/10.1038/s41598-019-53255-6} {\bibfield  {journal} {\bibinfo
  {journal} {Scientific Reports}\ }\textbf {\bibinfo {volume} {9}},\ \bibinfo
  {pages} {17474} (\bibinfo {year} {2019})}\BibitemShut {NoStop}%
\bibitem [{\citenamefont {Zvezdin}\ and\ \citenamefont
  {Kotov}(1997)}]{zvezdin1997modern}%
  \BibitemOpen
  \bibfield  {author} {\bibinfo {author} {\bibfnamefont {A.}~\bibnamefont
  {Zvezdin}}\ and\ \bibinfo {author} {\bibfnamefont {V.}~\bibnamefont
  {Kotov}},\ }\href {https://books.google.nl/books?id=hQ7Xk7MToRoC} {\emph
  {\bibinfo {title} {Modern Magnetooptics and Magnetooptical Materials}}},\
  Condensed Matter Physics\ (\bibinfo  {publisher} {CRC Press},\ \bibinfo
  {year} {1997})\BibitemShut {NoStop}%
\bibitem [{\citenamefont {Kirilyuk}\ \emph {et~al.}(2010)\citenamefont
  {Kirilyuk}, \citenamefont {Kimel},\ and\ \citenamefont
  {Rasing}}]{RevModPhys.82.2731}%
  \BibitemOpen
  \bibfield  {author} {\bibinfo {author} {\bibfnamefont {A.}~\bibnamefont
  {Kirilyuk}}, \bibinfo {author} {\bibfnamefont {A.~V.}\ \bibnamefont
  {Kimel}},\ and\ \bibinfo {author} {\bibfnamefont {T.}~\bibnamefont
  {Rasing}},\ }\bibfield  {title} {\bibinfo {title} {Ultrafast optical
  manipulation of magnetic order},\ }\href
  {https://doi.org/10.1103/RevModPhys.82.2731} {\bibfield  {journal} {\bibinfo
  {journal} {Rev. Mod. Phys.}\ }\textbf {\bibinfo {volume} {82}},\ \bibinfo
  {pages} {2731} (\bibinfo {year} {2010})}\BibitemShut {NoStop}%
\bibitem [{\citenamefont {Landau}\ and\ \citenamefont
  {Lifshitz}(1963)}]{LandauLifshitz}%
  \BibitemOpen
  \bibfield  {author} {\bibinfo {author} {\bibfnamefont {L.}~\bibnamefont
  {Landau}}\ and\ \bibinfo {author} {\bibfnamefont {E.}~\bibnamefont
  {Lifshitz}},\ }\href@noop {} {\emph {\bibinfo {title} {Course of Theoretical
  Physics: Vol. 8: Electrodynamics of Continous Media}}},\ \bibinfo {edition}
  {2nd}\ ed.\ (\bibinfo  {publisher} {Oxford: Reed Educational and Professional
  Publishing Ltd.},\ \bibinfo {year} {1963})\BibitemShut {NoStop}%
\bibitem [{\citenamefont {Pitaevskii}(1961)}]{pitaevskii1961electric}%
  \BibitemOpen
  \bibfield  {author} {\bibinfo {author} {\bibfnamefont {L.}~\bibnamefont
  {Pitaevskii}},\ }\bibfield  {title} {\bibinfo {title} {Electric forces in a
  transparent dispersive medium},\ }\href@noop {} {\bibfield  {journal}
  {\bibinfo  {journal} {Sov. Phys. JETP}\ }\textbf {\bibinfo {volume} {12}},\
  \bibinfo {pages} {1008} (\bibinfo {year} {1961})}\BibitemShut {NoStop}%
\bibitem [{\citenamefont {Pershan}(1963)}]{PhysRev.130.919}%
  \BibitemOpen
  \bibfield  {author} {\bibinfo {author} {\bibfnamefont {P.~S.}\ \bibnamefont
  {Pershan}},\ }\bibfield  {title} {\bibinfo {title} {Nonlinear optical
  properties of solids: Energy considerations},\ }\href
  {https://doi.org/10.1103/PhysRev.130.919} {\bibfield  {journal} {\bibinfo
  {journal} {Phys. Rev.}\ }\textbf {\bibinfo {volume} {130}},\ \bibinfo {pages}
  {919} (\bibinfo {year} {1963})}\BibitemShut {NoStop}%
\bibitem [{\citenamefont {Neumann}(1885)}]{Neumann}%
  \BibitemOpen
  \bibfield  {author} {\bibinfo {author} {\bibfnamefont {F.}~\bibnamefont
  {Neumann}},\ }\href@noop {} {\emph {\bibinfo {title} {Vorlesungen \"{u}ber
  die Theorie der Elastizit\"{a}t der festen K\"{o}rper und des
  Licht\"{a}thers}}}\ (\bibinfo  {publisher} {B. G. Teubner-Verlag},\ \bibinfo
  {address} {Leipzig},\ \bibinfo {year} {1885})\BibitemShut {NoStop}%
\bibitem [{\citenamefont {Pershan}\ \emph {et~al.}(1966)\citenamefont
  {Pershan}, \citenamefont {van~der Ziel},\ and\ \citenamefont
  {Malmstrom}}]{PhysRev.143.574}%
  \BibitemOpen
  \bibfield  {author} {\bibinfo {author} {\bibfnamefont {P.~S.}\ \bibnamefont
  {Pershan}}, \bibinfo {author} {\bibfnamefont {J.~P.}\ \bibnamefont {van~der
  Ziel}},\ and\ \bibinfo {author} {\bibfnamefont {L.~D.}\ \bibnamefont
  {Malmstrom}},\ }\bibfield  {title} {\bibinfo {title} {Theoretical discussion
  of the inverse {Faraday} effect, raman scattering, and related phenomena},\
  }\href {https://doi.org/10.1103/PhysRev.143.574} {\bibfield  {journal}
  {\bibinfo  {journal} {Phys. Rev.}\ }\textbf {\bibinfo {volume} {143}},\
  \bibinfo {pages} {574} (\bibinfo {year} {1966})}\BibitemShut {NoStop}%
\bibitem [{\citenamefont {Birss}(1964)}]{birss1964symmetry}%
  \BibitemOpen
  \bibfield  {author} {\bibinfo {author} {\bibfnamefont {R.}~\bibnamefont
  {Birss}},\ }\href {https://books.google.nl/books?id=Iz5RAAAAMAAJ} {\emph
  {\bibinfo {title} {Symmetry and magnetism}}},\ Selected topics in solid state
  physics\ (\bibinfo  {publisher} {North-Holland Pub. Co.},\ \bibinfo {year}
  {1964})\BibitemShut {NoStop}%
\bibitem [{\citenamefont {Kimel}\ \emph {et~al.}(2005)\citenamefont {Kimel},
  \citenamefont {Kirilyuk}, \citenamefont {Usachev}, \citenamefont {Pisarev},
  \citenamefont {Balbashov},\ and\ \citenamefont {Rasing}}]{Nat435Kimel}%
  \BibitemOpen
  \bibfield  {author} {\bibinfo {author} {\bibfnamefont {A.}~\bibnamefont
  {Kimel}}, \bibinfo {author} {\bibfnamefont {A.}~\bibnamefont {Kirilyuk}},
  \bibinfo {author} {\bibfnamefont {P.}~\bibnamefont {Usachev}}, \bibinfo
  {author} {\bibfnamefont {R.}~\bibnamefont {Pisarev}}, \bibinfo {author}
  {\bibfnamefont {A.}~\bibnamefont {Balbashov}},\ and\ \bibinfo {author}
  {\bibfnamefont {T.}~\bibnamefont {Rasing}},\ }\bibfield  {title} {\bibinfo
  {title} {Ultrafast non-thermal control of magnetization by instantaneous
  photomagnetic pulses},\ }\href {https://doi.org/10.1038/nature03564}
  {\bibfield  {journal} {\bibinfo  {journal} {Nature}\ }\textbf {\bibinfo
  {volume} {435}},\ \bibinfo {pages} {655} (\bibinfo {year}
  {2005})}\BibitemShut {NoStop}%
\bibitem [{\citenamefont {van~der Ziel}\ \emph {et~al.}(1965)\citenamefont
  {van~der Ziel}, \citenamefont {Pershan},\ and\ \citenamefont
  {Malmstrom}}]{PhysRevLett.15.190}%
  \BibitemOpen
  \bibfield  {author} {\bibinfo {author} {\bibfnamefont {J.~P.}\ \bibnamefont
  {van~der Ziel}}, \bibinfo {author} {\bibfnamefont {P.~S.}\ \bibnamefont
  {Pershan}},\ and\ \bibinfo {author} {\bibfnamefont {L.~D.}\ \bibnamefont
  {Malmstrom}},\ }\bibfield  {title} {\bibinfo {title} {Optically-induced
  magnetization resulting from the inverse {Faraday} effect},\ }\href
  {https://doi.org/10.1103/PhysRevLett.15.190} {\bibfield  {journal} {\bibinfo
  {journal} {Phys. Rev. Lett.}\ }\textbf {\bibinfo {volume} {15}},\ \bibinfo
  {pages} {190} (\bibinfo {year} {1965})}\BibitemShut {NoStop}%
\bibitem [{\citenamefont {Zvezdin}\ \emph {et~al.}(2020)\citenamefont
  {Zvezdin}, \citenamefont {Kimel}, \citenamefont {Plokhov},\ and\
  \citenamefont {Zvezdin}}]{Zvezdin2020}%
  \BibitemOpen
  \bibfield  {author} {\bibinfo {author} {\bibfnamefont {A.~K.}\ \bibnamefont
  {Zvezdin}}, \bibinfo {author} {\bibfnamefont {A.~V.}\ \bibnamefont {Kimel}},
  \bibinfo {author} {\bibfnamefont {D.~I.}\ \bibnamefont {Plokhov}},\ and\
  \bibinfo {author} {\bibfnamefont {K.~A.}\ \bibnamefont {Zvezdin}},\
  }\bibfield  {title} {\bibinfo {title} {Ultrafast spin dynamics in the iron
  borate easy-plane weak ferromagnet},\ }\href
  {https://doi.org/10.1134/S1063776120070195} {\bibfield  {journal} {\bibinfo
  {journal} {Journal of Experimental and Theoretical Physics}\ }\textbf
  {\bibinfo {volume} {131}},\ \bibinfo {pages} {130} (\bibinfo {year}
  {2020})}\BibitemShut {NoStop}%
\bibitem [{\citenamefont {Yan}\ and\ \citenamefont
  {Nelson}(1987)}]{doi:10.1063/1.453733}%
  \BibitemOpen
  \bibfield  {author} {\bibinfo {author} {\bibfnamefont {Y.}~\bibnamefont
  {Yan}}\ and\ \bibinfo {author} {\bibfnamefont {K.~A.}\ \bibnamefont
  {Nelson}},\ }\bibfield  {title} {\bibinfo {title} {Impulsive stimulated light
  scattering. i. general theory},\ }\href {https://doi.org/10.1063/1.453733}
  {\bibfield  {journal} {\bibinfo  {journal} {The Journal of Chemical Physics}\
  }\textbf {\bibinfo {volume} {87}},\ \bibinfo {pages} {6240} (\bibinfo {year}
  {1987})}\BibitemShut {NoStop}%
\bibitem [{\citenamefont {Kalashnikova}\ \emph {et~al.}(2007)\citenamefont
  {Kalashnikova}, \citenamefont {Kimel}, \citenamefont {Pisarev}, \citenamefont
  {Gridnev}, \citenamefont {Kirilyuk},\ and\ \citenamefont
  {Rasing}}]{PhysRevLett.99.167205}%
  \BibitemOpen
  \bibfield  {author} {\bibinfo {author} {\bibfnamefont {A.~M.}\ \bibnamefont
  {Kalashnikova}}, \bibinfo {author} {\bibfnamefont {A.~V.}\ \bibnamefont
  {Kimel}}, \bibinfo {author} {\bibfnamefont {R.~V.}\ \bibnamefont {Pisarev}},
  \bibinfo {author} {\bibfnamefont {V.~N.}\ \bibnamefont {Gridnev}}, \bibinfo
  {author} {\bibfnamefont {A.}~\bibnamefont {Kirilyuk}},\ and\ \bibinfo
  {author} {\bibfnamefont {T.}~\bibnamefont {Rasing}},\ }\bibfield  {title}
  {\bibinfo {title} {Impulsive generation of coherent magnons by linearly
  polarized light in the easy-plane antiferromagnet {${\mathrm{FeBO}}_{3}$}},\
  }\href {https://doi.org/10.1103/PhysRevLett.99.167205} {\bibfield  {journal}
  {\bibinfo  {journal} {Phys. Rev. Lett.}\ }\textbf {\bibinfo {volume} {99}},\
  \bibinfo {pages} {167205} (\bibinfo {year} {2007})}\BibitemShut {NoStop}%
\bibitem [{\citenamefont {Gridnev}(2008)}]{PhysRevB.77.094426}%
  \BibitemOpen
  \bibfield  {author} {\bibinfo {author} {\bibfnamefont {V.~N.}\ \bibnamefont
  {Gridnev}},\ }\bibfield  {title} {\bibinfo {title} {Phenomenological theory
  for coherent magnon generation through impulsive stimulated {Raman}
  scattering},\ }\href {https://doi.org/10.1103/PhysRevB.77.094426} {\bibfield
  {journal} {\bibinfo  {journal} {Phys. Rev. B}\ }\textbf {\bibinfo {volume}
  {77}},\ \bibinfo {pages} {094426} (\bibinfo {year} {2008})}\BibitemShut
  {NoStop}%
\bibitem [{\citenamefont {Kalashnikova}\ \emph {et~al.}(2008)\citenamefont
  {Kalashnikova}, \citenamefont {Kimel}, \citenamefont {Pisarev}, \citenamefont
  {Gridnev}, \citenamefont {Usachev}, \citenamefont {Kirilyuk},\ and\
  \citenamefont {Rasing}}]{PhysRevB.78.104301}%
  \BibitemOpen
  \bibfield  {author} {\bibinfo {author} {\bibfnamefont {A.~M.}\ \bibnamefont
  {Kalashnikova}}, \bibinfo {author} {\bibfnamefont {A.~V.}\ \bibnamefont
  {Kimel}}, \bibinfo {author} {\bibfnamefont {R.~V.}\ \bibnamefont {Pisarev}},
  \bibinfo {author} {\bibfnamefont {V.~N.}\ \bibnamefont {Gridnev}}, \bibinfo
  {author} {\bibfnamefont {P.~A.}\ \bibnamefont {Usachev}}, \bibinfo {author}
  {\bibfnamefont {A.}~\bibnamefont {Kirilyuk}},\ and\ \bibinfo {author}
  {\bibfnamefont {T.}~\bibnamefont {Rasing}},\ }\bibfield  {title} {\bibinfo
  {title} {Impulsive excitation of coherent magnons and phonons by
  subpicosecond laser pulses in the weak ferromagnet {${\text{FeBO}}_{3}$}},\
  }\href {https://doi.org/10.1103/PhysRevB.78.104301} {\bibfield  {journal}
  {\bibinfo  {journal} {Phys. Rev. B}\ }\textbf {\bibinfo {volume} {78}},\
  \bibinfo {pages} {104301} (\bibinfo {year} {2008})}\BibitemShut {NoStop}%
\bibitem [{\citenamefont {Juraschek}\ \emph {et~al.}(2020)\citenamefont
  {Juraschek}, \citenamefont {Narang},\ and\ \citenamefont
  {Spaldin}}]{PhysRevResearch.2.043035}%
  \BibitemOpen
  \bibfield  {author} {\bibinfo {author} {\bibfnamefont {D.~M.}\ \bibnamefont
  {Juraschek}}, \bibinfo {author} {\bibfnamefont {P.}~\bibnamefont {Narang}},\
  and\ \bibinfo {author} {\bibfnamefont {N.~A.}\ \bibnamefont {Spaldin}},\
  }\bibfield  {title} {\bibinfo {title} {Phono-magnetic analogs to
  opto-magnetic effects},\ }\href
  {https://doi.org/10.1103/PhysRevResearch.2.043035} {\bibfield  {journal}
  {\bibinfo  {journal} {Phys. Rev. Research}\ }\textbf {\bibinfo {volume}
  {2}},\ \bibinfo {pages} {043035} (\bibinfo {year} {2020})}\BibitemShut
  {NoStop}%
\bibitem [{\citenamefont {Blank}\ \emph {et~al.}(2022)\citenamefont {Blank},
  \citenamefont {Grishunin}, \citenamefont {Ivanov}, \citenamefont
  {Mashkovich}, \citenamefont {Afanasiev},\ and\ \citenamefont
  {Kimel}}]{blank2022empowering}%
  \BibitemOpen
  \bibfield  {author} {\bibinfo {author} {\bibfnamefont {T.~G.~H.}\
  \bibnamefont {Blank}}, \bibinfo {author} {\bibfnamefont {K.~A.}\ \bibnamefont
  {Grishunin}}, \bibinfo {author} {\bibfnamefont {B.~A.}\ \bibnamefont
  {Ivanov}}, \bibinfo {author} {\bibfnamefont {E.~A.}\ \bibnamefont
  {Mashkovich}}, \bibinfo {author} {\bibfnamefont {D.}~\bibnamefont
  {Afanasiev}},\ and\ \bibinfo {author} {\bibfnamefont {A.}~\bibnamefont
  {Kimel}},\ }\bibfield  {title} {\bibinfo {title} {Empowering control of
  antiferromagnets by {THz}-induced spin coherence},\ }\href
  {https://arxiv.org/abs/2212.09532} {\bibfield  {journal} {\bibinfo  {journal}
  {arXiv preprint arXiv:2212.09532}\ } (\bibinfo {year} {2022})}\BibitemShut
  {NoStop}%
\bibitem [{\citenamefont {Gruber}\ \emph {et~al.}(1989)\citenamefont {Gruber},
  \citenamefont {Hills}, \citenamefont {Macfarlane}, \citenamefont {Morrison},
  \citenamefont {Turner}, \citenamefont {Quarles}, \citenamefont {Kintz},\ and\
  \citenamefont {Esterowitz}}]{PhysRevB.40.9464}%
  \BibitemOpen
  \bibfield  {author} {\bibinfo {author} {\bibfnamefont {J.~B.}\ \bibnamefont
  {Gruber}}, \bibinfo {author} {\bibfnamefont {M.~E.}\ \bibnamefont {Hills}},
  \bibinfo {author} {\bibfnamefont {R.~M.}\ \bibnamefont {Macfarlane}},
  \bibinfo {author} {\bibfnamefont {C.~A.}\ \bibnamefont {Morrison}}, \bibinfo
  {author} {\bibfnamefont {G.~A.}\ \bibnamefont {Turner}}, \bibinfo {author}
  {\bibfnamefont {G.~J.}\ \bibnamefont {Quarles}}, \bibinfo {author}
  {\bibfnamefont {G.~J.}\ \bibnamefont {Kintz}},\ and\ \bibinfo {author}
  {\bibfnamefont {L.}~\bibnamefont {Esterowitz}},\ }\bibfield  {title}
  {\bibinfo {title} {Spectra and energy levels of{
  ${\mathrm{Tm}}^{3+}$:${\mathrm{Y}}_{3}$${\mathrm{Al}}_{5}$${\mathrm{O}}_{12}$}},\
  }\href {https://doi.org/10.1103/PhysRevB.40.9464} {\bibfield  {journal}
  {\bibinfo  {journal} {Phys. Rev. B}\ }\textbf {\bibinfo {volume} {40}},\
  \bibinfo {pages} {9464} (\bibinfo {year} {1989})}\BibitemShut {NoStop}%
\bibitem [{\citenamefont {Tiseanu}\ \emph {et~al.}(1995)\citenamefont
  {Tiseanu}, \citenamefont {Lupei},\ and\ \citenamefont
  {Lupei}}]{Tiseanu_1995}%
  \BibitemOpen
  \bibfield  {author} {\bibinfo {author} {\bibfnamefont {C.}~\bibnamefont
  {Tiseanu}}, \bibinfo {author} {\bibfnamefont {A.}~\bibnamefont {Lupei}},\
  and\ \bibinfo {author} {\bibfnamefont {V.}~\bibnamefont {Lupei}},\ }\bibfield
   {title} {\bibinfo {title} {Energy levels of {Tm$^{3+}$} in yttrium aluminium
  garnet},\ }\href {https://doi.org/10.1088/0953-8984/7/44/016} {\bibfield
  {journal} {\bibinfo  {journal} {J. Phys.: Condensed Matter}\ }\textbf
  {\bibinfo {volume} {7}},\ \bibinfo {pages} {8477} (\bibinfo {year}
  {1995})}\BibitemShut {NoStop}%
\bibitem [{\citenamefont {Ju}\ \emph {et~al.}(2019)\citenamefont {Ju},
  \citenamefont {Zhong}, \citenamefont {Lu},\ and\ \citenamefont
  {Yeung}}]{doi:10.1021/acs.inorgchem.8b02009}%
  \BibitemOpen
  \bibfield  {author} {\bibinfo {author} {\bibfnamefont {M.}~\bibnamefont
  {Ju}}, \bibinfo {author} {\bibfnamefont {M.}~\bibnamefont {Zhong}}, \bibinfo
  {author} {\bibfnamefont {C.}~\bibnamefont {Lu}},\ and\ \bibinfo {author}
  {\bibfnamefont {Y.-y.}\ \bibnamefont {Yeung}},\ }\bibfield  {title} {\bibinfo
  {title} {Deciphering the microstructure and energy-level splitting of
  {Tm$^{3+}$}-doped yttrium aluminum garnet},\ }\href
  {https://doi.org/10.1021/acs.inorgchem.8b02009} {\bibfield  {journal}
  {\bibinfo  {journal} {Inorganic Chemistry}\ }\textbf {\bibinfo {volume}
  {58}},\ \bibinfo {pages} {1058} (\bibinfo {year} {2019})},\ \bibinfo {note}
  {pMID: 30216052}\BibitemShut {NoStop}%
\bibitem [{\citenamefont {Seki}\ \emph {et~al.}(2015)\citenamefont {Seki},
  \citenamefont {Uchida}, \citenamefont {Kikkawa}, \citenamefont {Qiu},
  \citenamefont {Saitoh},\ and\ \citenamefont
  {Takanashi}}]{doi:10.1063/1.4929691}%
  \BibitemOpen
  \bibfield  {author} {\bibinfo {author} {\bibfnamefont {T.}~\bibnamefont
  {Seki}}, \bibinfo {author} {\bibfnamefont {K.-i.}\ \bibnamefont {Uchida}},
  \bibinfo {author} {\bibfnamefont {T.}~\bibnamefont {Kikkawa}}, \bibinfo
  {author} {\bibfnamefont {Z.}~\bibnamefont {Qiu}}, \bibinfo {author}
  {\bibfnamefont {E.}~\bibnamefont {Saitoh}},\ and\ \bibinfo {author}
  {\bibfnamefont {K.}~\bibnamefont {Takanashi}},\ }\bibfield  {title} {\bibinfo
  {title} {Observation of inverse spin {Hall} effect in ferromagnetic {FePt}
  alloys using spin {Seebeck} effect},\ }\href
  {https://doi.org/10.1063/1.4929691} {\bibfield  {journal} {\bibinfo
  {journal} {Applied Physics Letters}\ }\textbf {\bibinfo {volume} {107}},\
  \bibinfo {pages} {092401} (\bibinfo {year} {2015})}\BibitemShut {NoStop}%
\end{thebibliography}%
\end{document}


\title{Supplemental Material for ``Effective rectification of THz electromagnetic fields in a ferrimagnetic iron garnet''}
\author{T.G.H. Blank}
\affiliation{Radboud University, Institute for Molecules and Materials, 6525 AJ Nijmegen, the Netherlands.}
\affiliation{Department of Applied Physics, Eindhoven University of Technology, P.O. Box 513, Eindhoven 5600 MB, the Netherlands.}
\author{E.A. Mashkovich}
\affiliation{University of Cologne, Institute of Physics II, Cologne D-50937, Germany.}
\author{K.A. Grishunin}
\affiliation{Radboud University, Institute for Molecules and Materials, 6525 AJ Nijmegen, the Netherlands.}
\author{C. Schippers}
\affiliation{Department of Applied Physics, Eindhoven University of Technology, P.O. Box 513, Eindhoven 5600 MB, the Netherlands.}
\author{M.V. Logunov}
\affiliation{Kotel'nikov Institute of Radioengineering and Electronics, 125009 Moscow, Russia.}
\author{B. Koopmans}
\affiliation{Department of Applied Physics, Eindhoven University of Technology, P.O. Box 513, Eindhoven 5600 MB, the Netherlands.}
\author{A.K. Zvezdin}
\affiliation{Prokhorov General Physics Institute of the Russian Academy of Sciences, 119991 Moscow, Russia.}
\author{A.V. Kimel}
\affiliation{Radboud University, Institute for Molecules and Materials, 6525 AJ Nijmegen, the Netherlands.}

\date{\small{\today}}

\maketitle
\section*{XRD Analysis}
\begin{figure}[h!]
    \centering
    \includegraphics[scale = 0.95]{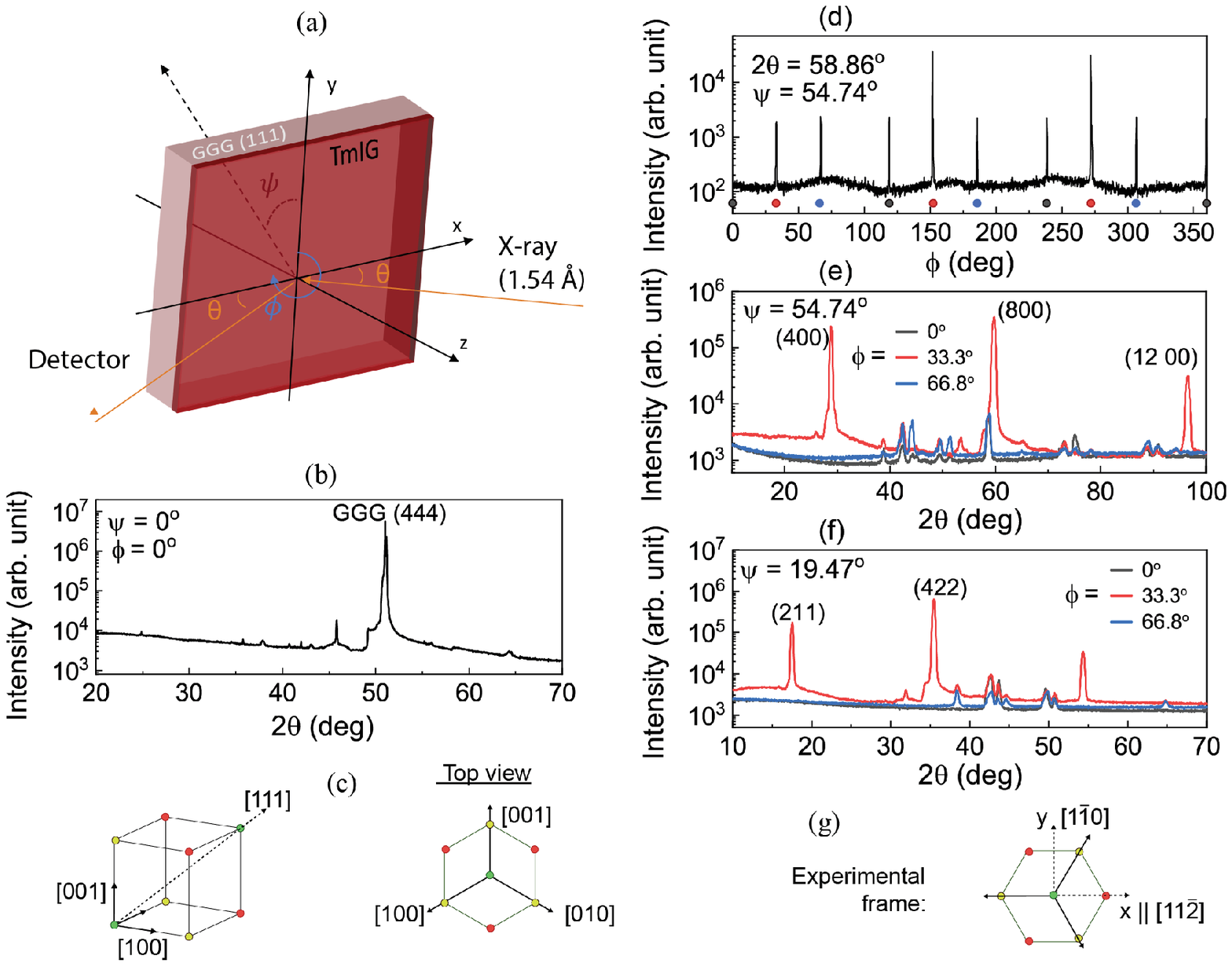}
    \caption{\textit{(a) Schematic illustration of the angles involved in the XRD analysis of the sample, where $x$, $y$ and $z$ coincide with the experimental axes from Fig.~1(a) of the main article. (b) XRD data for the sample plane, which confirms the $[111]$ orientation of the sample \cite{doi:10.1063/1.4929691}. (c) Illustration of the cubic crystal axes and the two-dimensional projection seen from the $[111]$ direction. (d) Scattering intensity at $2\theta = 58.86^\circ$ ($(800)$) for different sample rotations $\phi$, where the sample tilt is set at $\psi = 54.74^\circ$ which is equal to the angle between the $[111]$ and $[100]$ axes. This data suggests three possible candidates $\phi = 0^\circ, 33.3^\circ, 66.6^\circ$ which could coincide with a $(100)$ plane. (e) Angular scattering intensity measurements for a tilting angle of $\psi = 54.74^\circ$ and for the three different candidate angles $\phi$. The data at $\phi = 33.3^\circ$ shows what is typical for a $(100)$ plane. In (f), we perform similar measurements, but for a tilting angle $\psi = 19.47^\circ $, corresponding to the angle between the $[111]$ and $[112]$ axes. The result confirms that $\phi = 33.3^\circ$ corresponds to a $(211)$ plane. (g) The results allowed us to conclude that the projections of the cubic axes are rotated at an angle of $33.3^\circ$ compared to the drawing in (c), meaning that the experimental $x$ and $y$ axis correspond to the $[11\overline{2}]$ and $[1\overline{1}0]$ axes, respectively.}}
    \label{fig:XRD}
\end{figure}

We performed X-ray diffraction (XRD) analysis on the sample to obtain the in-plane crystallographic orientation. Fig.~\ref{fig:XRD}(a) depicts the experimental setup. The scattering plane of the sample can be tilted at an angle $\psi$, and can be rotated around its normal axis by an angle $\phi$. Because cubic crystals are self-dual, we can identify the Miller planes $(hkl)$ to be simply the planes perpendicular to the vector $[hkl]$. 

The data for $\phi = 0^\circ$, and $\psi = 0^\circ$ confirms the $[111]$ orientation of the sample \cite{doi:10.1063/1.4929691}. From the illustrations in Fig.~\ref{fig:XRD}(c), it can be seen that after tilting the sample at $\psi = 54.74^\circ$ (angle between $[111]$ and $[100]$), it should be possible to let the scattering plane coincide with one of the base vector planes $(100)$, $(010)$ or $(001)$ by rotating the sample over a certain angle $\phi$. Therefore, we put $\psi = 54.74^\circ$ and set the radiation and detection angle at $2\theta = 58.86^\circ$ corresponding to a $(800)$ peak to recognize a $(100)$ plane. By scanning the rotation angle $\phi$, we obtained three candidates at $\phi = 0^\circ, 33.3^\circ, 66.6^\circ$ (see Fig.~\ref{fig:XRD}(d)). Due to the three-fold symmetric $[111]$ axis, the peaks at $\phi + n\cdot 2\pi/3$ are equivalent. The data in Fig.~\ref{fig:XRD}(e) establishes that only $\phi= 33.3^\circ$ (and equivalent angles) coincides with a $(100)$ plane. For additional confirmation, we note that by tilting the sample at $\psi = 19.47^\circ$, the same rotation $\phi= 33.3^\circ $ should correspond to a $(211)$ plane. This is confirmed in Fig.~\ref{fig:XRD}(f). Therefore, we conclude that the crystallographic axes in our experiment are oriented like the dotted arrows in Fig.~\ref{fig:XRD}(g).

Note that although the peaks at $\phi + n\cdot 2\pi/3$ with $n = 0, 1, 2$ should be equivalent, it turns out that the equivalent axes for $\phi = 33.3^\circ$ were rather $\phi = 150^\circ$ and $\phi = 270^\circ$, meaning that the first rotation is slightly shifted. Similar results were obtained in Ref.~\cite{Ciubotariu2019} and were explained as a result of strain between the thulium iron garnet film and the Gd$_3$Fe$_5$O$_{12}$ substrate.
\newpage

\bibliography{references}